\documentclass[11pt]{article}
\usepackage{biometrics_woojung}
\usepackage{enumitem}
\usepackage{natbib}
\usepackage{tikz}
\usetikzlibrary{arrows.meta, positioning}
\usepackage{changes}

\title{A Causal Framework for Quantile Residual Lifetime}

\author[1]{Taekwon Hong}
\author[2]{Woojung Bae}
\author[3]{Sang Kyu Lee}
\author[4]{Dongrak Choi}
\author[5]{Jong-Hyeon Jeong}

\affil[1]{Division of Biometrics VII, Office of Biostatistics, Office of Translational Sciences, Center for Drug Evaluation and Research, U.S. Food and Drug Administration, Silver Spring, Maryland 20993}
\affil[2]{Division of Biostatistics, Office of Biostatistics and Pharmacovigilance, Center for Biologics Evaluation and Research, U.S. Food and Drug Administration, Silver Spring, Maryland 20993}
\affil[3]{Department of Applied Statistics, Konkuk University, Seoul 05029}
\affil[4]{Department of Biostatistics and Bioinformatics, Duke University, Durham, North Carolina 27705}
\affil[5]{Biometric Research Program, Division of Cancer Treatment and Diagnosis, National Cancer Institute, Bethesda, Maryland 20892}

\makeatletter

\begin{document}

\maketitle

\setlength{\parskip}{2pt}
\setlength{\abovedisplayskip}{2pt}
\setlength{\belowdisplayskip}{2pt}
\setlength{\abovedisplayshortskip}{2pt}
\setlength{\belowdisplayshortskip}{2pt}

\begin{abstract}
    {
    Estimating prognosis conditional on surviving an initial high-risk period is crucial in clinical research. Yet, standard metrics such as hazard ratios are often difficult to interpret, while mean-based summaries are sensitive to outliers and censoring. We propose a formal causal framework for estimating quantiles of residual lifetime among individuals surviving to a landmark time $t_0$. Our primary estimand, the ``Observed Survivor Quantile Contrast'' (OSQC), targets pragmatic prognostic differences within the observed survivor population. To estimate the OSQC, we develop a doubly robust estimator that combines propensity scores, outcome regression, and inverse probability of censoring weights, ensuring consistency under confounding and informative censoring provided that the censoring model is correctly specified and at least one additional nuisance model is correctly specified. Recognizing that the OSQC conflates causal efficacy and compositional selection, we also introduce a reweighting-based supplementary estimator for the ``Principal Survivor Quantile Contrast'' (PSQC) to disentangle these mechanisms under stronger assumptions. Extensive simulations demonstrate the robustness of the proposed estimators and clarify the role of post-treatment selection. We illustrate the framework using data from the SUPPORT study to assess the impact of right heart catheterization on residual lifetime among intensive care unit survivors, and from the NSABP B-14 trial to examine post-surgical prognosis under adjuvant tamoxifen therapy across multiple landmark times.

    \keywords{Causal inference; Quantile residual life; Doubly robust estimation; Principal stratification}
    }
\end{abstract}

\newpage
\section{Introduction}
\label{sec:intro}

In many clinical and epidemiological studies, investigators are interested not only in overall survival differences between treatment groups, but also in prognosis conditional on surviving an initial high-risk period. Such questions naturally arise in settings with acute early mortality, delayed treatment effects or clinical stabilization phases where the interpretation of marginal survival contrasts may be limited. For example, among patients who survive an initial hospitalization or intensive care phase, clinicians may wish to understand how treatment affects subsequent survival prospects rather than cumulative mortality from baseline \citep{zabor2013dynamic,hieke2015conditional}.

Landmark analysis constitutes a widely adopted framework for addressing this objective, conditioning on survival up to a prespecified time $t_0$ to evaluate subsequent outcomes \citep[][]{dafni2011landmark}. While existing approaches in this setting predominantly focus on mean residual life \citep[e.g.,][]{oakes1990note, chen2005semiparametric, chan2012proportional}, they often rely on restrictive proportionality assumptions. Furthermore, mean-based summaries are inherently sensitive to skewness and extreme observations \citep{jeong2014statistical}. These limitations motivate a shift toward distributional summaries of remaining survival time. The quantile residual lifetime approach offers robust, directly interpretable prognostic measures that allow for the investigation of clinically relevant regions of the distribution---such as lower-tail quantiles, which characterize survival prospects for high-risk subpopulations \citep[e.g.,][]{jung2009regression,kim2012censored,lin2019quantile}.

Despite their appeal, formal causal frameworks for quantiles of residual lifetime remain underdeveloped. Conditioning on survival to $t_0$ induces post-treatment selection, making causal estimands inherently nontrivial to define even at a conceptual level \citep{rosenbaum1984consequences, hernan2004structural, rubin2006causal, stensrud2022separable}. Moreover, informative censoring is common in survival settings and may be driven by baseline prognostic factors and treatment itself, further complicating identification and estimation in practice \citep{robins1986new, robins2000correcting}. While quantile regression methods for censored data are well established in descriptive contexts \citep[e.g.,][]{koenker2001reappraising,portnoy2003censored,peng2008survival,wei2021estimation}, comparatively few tools exist for causal quantile estimation of residual lifetime under realistic longitudinal data structures.

These conceptual challenges have motivated renewed attention to how causal comparisons are defined in time-to-event settings. Recent works \citep{martinussen2022causality,beyersmann2025hazards} emphasize that causal interpretation hinges on how survivor populations are defined over time. In particular, contrasts formulated on the probability scale admit a clear causal interpretation under randomization because they correspond to well-defined comparisons between counterfactual outcome distributions. Building on these insights, we develop a causal framework for estimating quantiles of residual lifetime conditional on survival to a landmark time $t_0$. Our primary estimand is the contrast in residual-life quantiles between treatment arms among individuals who survive to a prespecified landmark time $t_0$ under each treatment regime, which we refer to as the \emph{Observed Survivor Quantile Contrast} (OSQC). By indexing the estimand explicitly through the landmark time $t_0$ and the quantile level $\tau$, the proposed framework clarifies the target survivor population and the aspect of the outcome distribution being contrasted, enabling causal interpretation on the probability scale while accounting for post-treatment selection and censoring. Furthermore, although we recognize that the principal stratification framework \citep{frangakis2002principal, martinussen2020subtleties} relies on strong and untestable assumptions \citep{pearl2011principal, vanderweele2011principal, stensrud2022translating}, it remains valuable for disentangling compositional effects. Therefore, rather than using it for primary identification, we adapt it into a supplementary analysis that targets such estimands through simple additional weighting, offering researchers a flexible inferential tool.

The remainder of the paper is organized as follows.
Section~\ref{sec:causal_residual} introduces the causal framework for quantile residual lifetime, defines the target estimands and states the identifying assumptions.
Section~\ref{sec:estimation} presents the proposed estimation procedures, including the inversely weighted and doubly robust (DR) estimators and discusses statistical inference.
Section~\ref{sec:supplement} develops a supplementary analysis based on principal stratification to disentangle prognostic and compositional effects.
Section~\ref{sec:simulation} reports results from a comprehensive simulation study assessing finite-sample performance and robustness, while Section~\ref{sec:application} demonstrates the practical implementation of the proposed estimators.
Finally, Section~\ref{sec:discussion} concludes with a discussion of the methodological implications, limitations and potential extensions of this work.

\section{Causal framework for quantile residual lifetime}
\label{sec:causal_residual}

\subsection{Setup and target estimand}
\label{subsec:setup}

Let $A \in \{0,1\}$ denote a baseline treatment, and let $X \in \mathcal{X} \subset \mathbb{R}^p$ collect baseline covariates. We are interested in a nonnegative event time $T$ (e.g., failure or death time) and a right-censoring time $C$. Define the observed follow-up time and event indicator by
\[
Y = \min(T,C), \qquad \Delta = I(T \le C).
\]

We fix a clinically meaningful landmark time $t_0 > 0$. The primary outcome of interest is the \emph{residual lifetime} after $t_0$. For each treatment level $a \in \{0,1\}$, let $T_a$ denote the potential event time under treatment $a$. The potential residual lifetime is defined as
\[
R_a(t_0) = T_a - t_0 \quad \text{on the event } \{T_a > t_0\}.
\]

Our target is the distribution of $R_a(t_0)$ among the subpopulation of individuals who would survive to $t_0$ under treatment $a$. Formally, for $r \ge 0$, the cumulative distribution function (CDF) is
\begin{equation}
F_{R_a}(r; t_0) := P\bigl( T_a \le t_0+r \,\big|\, T_a>t_0 \bigr) = \frac{ P(t_0 < T_a \le t_0+r) }{ P(T_a > t_0) }.
\label{eq:FR_def}
\end{equation}
Correspondingly, the $\tau$-th quantile of the residual lifetime is defined as
\begin{equation}
q_a(\tau; t_0) = \inf\{r \ge 0 : F_{R_a}(r; t_0) \ge \tau\}, \qquad 0<\tau<1.
\label{eq:qa_def}
\end{equation}

We define the causal effect of treatment as the difference in residual lifetime quantiles between the two arms:
\begin{equation}
\delta(\tau; t_0) := q_1(\tau; t_0) - q_0(\tau; t_0).
\label{eq:delta_def}
\end{equation}
Positive values of $\delta(\tau; t_0)$ indicate that treatment $A=1$ extends the $\tau$-th quantile of remaining life compared to $A=0$. Because the conditioning event $T_a > t_0$ is post-treatment, $\delta(\tau; t_0)$ compares the prognosis of the potentially different sub-populations that survive under each arm. We refer to this quantity as the OSQC emphasizing that it targets the outcome distribution of the clinically observable survivor cohort rather than a fixed latent population.

In this work, we prioritize the OSQC because it answers the pragmatic clinical question: ``Given that a patient has survived to time $t_0$ under their assigned treatment, what is their expected remaining lifetime?'' While alternative estimands defined on latent principal strata can theoretically disentangle causal efficacy from compositional differences, they rely on stronger assumptions that may not be plausible in all settings. Therefore, we focus on the OSQC as the primary estimand and reserve the identification of latent-stratum effects for a supplementary analysis in Section~\ref{sec:supplement}.

\subsection{Assumptions}
The observed data consist of i.i.d.\ replicates \(O=(X,A,Y,\Delta)\), where \(Y=\min(T,C)\) and \(\Delta=I(T\le C)\). We write \(P\) for the true, unknown distribution of \(O\), and all expectations and probabilities are with respect to \(P\) unless stated otherwise. We now state the assumptions under which the causal distribution $F_{R_a}(r;t_0)$ is identified from the observed data $O$.

\begin{assumption}[Consistency]
\label{ass:consistency}
For each individual and $a\in\{0,1\}$, if $A=a$ then $T=T_a$ and $C=C_a$.
\end{assumption}

\begin{assumption}[Positivity]
\label{ass:positivity}
For almost every $x$ in the support of $X$,
\[
0 < e_a(x) := P(A=a \mid X=x) < 1.
\]
Furthermore, for all \(t\) in the support of \(Y\) and almost every \(x\),
\[
G_a(t\mid X=x):=P(C\ge t\mid A=a,X=x)>0.
\]
\end{assumption}

\begin{assumption}[Conditional exchangeability]
\label{ass:exchangeability}
For each $a \in \{0,1\}$,
\[
T_a \perp A \mid X.
\]
\end{assumption}

\begin{assumption}[Independent censoring given baseline covariates]
\label{ass:censoring}
For each \(a\in\{0,1\}\),
\[
\Pr(C_a \ge t \mid T_a \ge t, X) = \Pr(C_a \ge t \mid X)
\quad \text{for all } t.
\]
\end{assumption}

\begin{figure}[ht]
\centering
\begin{tikzpicture}[
    >=Stealth,
    node distance=1.8cm and 1.8cm,
    thick,
    state/.style={circle, draw, minimum size=0.9cm, align=center, inner sep=0pt},
    every edge/.style={draw, ->, shorten >=1pt}
]
\node[state] (X) {$X$};
\node[state, right=of X] (A) {$A$};
\node[state, above right=0.8cm and 1.2cm of A] (T) {$T$};
\node[state, below right=0.8cm and 1.2cm of A] (C) {$C$};

\draw[->] (X) -- (A);
\draw[->] (X) edge[bend left=25] (T);
\draw[->] (X) edge[bend right=25] (C);
\draw[->] (A) -- (T);
\draw[->] (A) -- (C);
\end{tikzpicture}
\caption{Causal DAG representing the data structure. 
\(X\): baseline covariates; \(A\): baseline treatment; \(T\): event time; \(C\): censoring time. Identification assumes no unmeasured confounding of \((A,T)\) given \(X\), and independent censoring given \((A,X)\).}
\label{fig:causal_dag}
\end{figure}
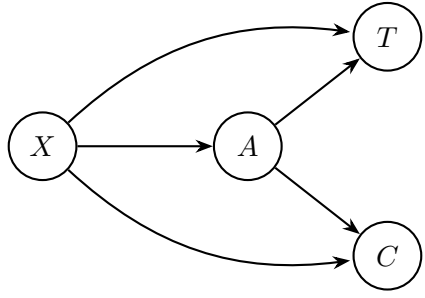

Assumption~\ref{ass:consistency} links the observed data to the counterfactuals implying that the treatment is well-defined and corresponds to the potential outcomes utilized in the analysis \citep{cole2009consistency}. Assumption~\ref{ass:positivity} ensures that individuals at every level of the covariates have a non-zero probability of receiving either treatment and remaining uncensored, thereby preventing structural violations that would render the effect non-identifiable \citep{westreich2010positivity}. Assumption~\ref{ass:exchangeability} is the standard ``no unmeasured confounding'' condition for point-exposure studies \citep{rosenbaum1983central}, positing that treatment assignment is effectively randomized within strata of baseline covariates. 
Finally, Assumption~\ref{ass:censoring} handles selection bias due to loss to follow-up. Unlike the stronger assumption of completely independent censoring required by the Kaplan-Meier (KM) estimator, this assumption allows the censoring mechanism to depend on baseline covariates \(X\) \citep{robins2000correcting}. Under Assumptions~\ref{ass:consistency} and \ref{ass:censoring}, $G_a(t\mid X) = \Pr(C \ge t \mid A=a, X, T \ge t)$, which justifies the inverse probability of censoring weighting (IPCW) representations. A Causal Directed Acyclic Graph (DAG) representing the data structure is provided in Figure \ref{fig:causal_dag}.

\subsection{Identification of the Causal Residual Lifetime Distribution}

We now show that, under Assumptions~\ref{ass:consistency}--\ref{ass:censoring}, $F_{R_a}(r;t_0)$ is identified as a functional of the observed data distribution. The key representation leverages inverse probability weighting (IPW) to adjust for treatment assignment and IPCW to account for loss to follow-up. 

\begin{theorem}[Identification of the residual-lifetime CDF]
\label{thm:identification}
Suppose Assumptions~\ref{ass:consistency}--\ref{ass:censoring} hold. Then, for each $a\in\{0,1\}$, $t_0>0$ and $r\ge 0$, the causal distribution $F_{R_a}(r;t_0)$ in \eqref{eq:FR_def} is identified and satisfies
\begin{equation}
F_{R_a}(r; t_0)
=
\frac{
E\!\left[
\dfrac{ I\bigl(A=a, \ t_0<Y\le t_0+r, \ \Delta=1\bigr) }
      { e_a(X)\, G_a(Y \mid X) }
\right]
}{
E\!\left[
\dfrac{ I\bigl(A=a, \ Y>t_0\bigr) }
      { e_a(X)\, G_a(t_0 \mid X) }
\right]
}.
\label{eq:FR_IPW_IPCW}
\end{equation}
In particular, the quantile $q_a(\tau;t_0)$ defined in \eqref{eq:qa_def} is identified for all $0<\tau<1$.
\end{theorem}

Theorem~\ref{thm:identification} provides a constructive identification formula that connects the counterfactual distribution of residual lifetime to the observed data distribution. The numerator in \eqref{eq:FR_IPW_IPCW} can be interpreted as a weighted expectation of the observed event process. The term $[e_a(X)]^{-1}$ adjusts for treatment assignment bias (confounding) creating a pseudo-population where treatment is independent of baseline covariates. Simultaneously, the term $[G_a(Y \mid X)]^{-1}$ adjusts for selection bias due to censoring and up-weighting individuals who remain under observation to represent those who were censored.

Crucially, this identification result relies on the weighting of the \emph{joint} event of survival to $t_0$ and failure by $t_0+r$. This structure allows us to estimate the conditional distribution $F_{R_a}$ without needing to model the full survival curve parametrically. The detailed proof of Theorem~\ref{thm:identification} is provided in the Supplementary Material.

\section{Estimation and Inference}
\label{sec:estimation}

Based on Theorem~\ref{thm:identification}, we now propose estimation procedures for the quantile $q_a(\tau; t_0)$. We begin with an IPW-IPCW-based (IW) estimator and then introduce a DR estimator to improve efficiency and reduce model dependence. Recall from definition \eqref{eq:qa_def} that the target parameter $\theta_a = q_a(\tau; t_0)$ satisfies the condition $F_{R_a}(\theta_a; t_0) = \tau$.
Substituting the identification result \eqref{eq:FR_IPW_IPCW} into the left-hand side, we require
\[
\frac{
E\!\left[ \dfrac{ I(A=a, \ t_0<Y\le t_0+\theta_a, \ \Delta=1) }{ e_a(X)\, G_a(Y \mid X) } \right]
}{
E\!\left[ \dfrac{ I(A=a, \ Y>t_0) }{ e_a(X)\, G_a(t_0 \mid X) } \right]
}
= \tau.
\]
Multiplying both sides by the denominator and rearranging terms, we obtain the following moment condition:
\begin{equation} \label{eq:fund}
E\!\left[
\frac{I(A=a)}{e_a(X)}
\left(
\frac{ I(t_0 < Y \le t_0 + \theta_a, \ \Delta=1) }{ G_a(Y \mid X) }
-
\tau
\frac{ I(Y > t_0) }{ G_a(t_0 \mid X) }
\right)
\right]
= 0.
\end{equation}
This population moment condition suggests that $q_a(\tau; t_0)$ can be estimated by finding the root of an empirical estimating equation.

\subsection{IW Estimator}

Let $\widehat{e}_a(X)$ and $\widehat{G}_a(t \mid X)$ denote consistent estimators for the propensity score and the censoring distribution, respectively. Based on the equation \eqref{eq:fund}, we define the estimated causal quantile $\widehat{q}_a(\tau; t_0)$ as the solution $\theta$ to the following estimating equation:
\begin{equation}
U_n^{IW}(\theta; \widehat{\eta}) 
= 
\frac{1}{n} \sum_{i=1}^n 
\frac{I(A_i=a)}{\widehat{e}_a(X_i)} 
\left[
\frac{ I(t_0 < Y_i \le t_0 + \theta, \ \Delta_i=1) }{ \widehat{G}_a(Y_i \mid X_i) }
- 
\tau 
\frac{ I(Y_i > t_0) }{ \widehat{G}_a(t_0 \mid X_i) }
\right]
\approx 0,
\label{eq:estimating_eq}
\end{equation}
where $\widehat{\eta} = (\widehat{e}_a, \widehat{G}_a)$ collects the nuisance parameters.

Since the indicator function $I(Y_i \le t_0 + \theta)$ is a step function in $\theta$, the estimating function in \eqref{eq:estimating_eq} is not continuous and may not equal zero exactly. However, note that it is a monotone non-decreasing function of $\theta$ (as $\theta$ increases, the first term in the bracket accumulates more positive weights) ensuring that the solution is unique. Following standard generalized estimating equation and quantile estimation literature, we define the estimator $\widehat{q}^{IW}_a(\tau; t_0)$ as the generalized inverse solution to equation \eqref{eq:estimating_eq}.

\subsection{Proposed DR Estimator}

To improve efficiency and robustness against model misspecification, we can augment the IW estimating equation \eqref{eq:estimating_eq} with an outcome regression model. Provided that the censoring mechanism is consistently estimated, this formulation yields a DR estimator \citep{chernozhukov2018double} that remains consistent if either the treatment assignment or the outcome model is correctly specified. Let $\mu_a(t \mid x) = P(T > t \mid A=a, X=x)$ denote the conditional survival probability given baseline covariates. We can estimate this quantity using a working outcome model such as a treatment-stratified Cox proportional hazards model.

The augmented estimating equation is constructed by subtracting the projection of the score onto the nuisance tangent space. The DR estimating function is defined as:
\begin{align}
U^{DR}_n(\theta) 
&= 
\frac{1}{n} \sum_{i=1}^n \Bigg\{
\underbrace{
\frac{I(A_i=a)}{\widehat{e}_a(X_i)} 
\left[
\frac{ I(t_0 < Y_i \le t_0 + \theta, \Delta_i=1) }{ \widehat{G}_a(Y_i \mid X_i) }
- 
\tau 
\frac{ I(Y_i > t_0) }{ \widehat{G}_a(t_0 \mid X_i) }
\right]
}_{\text{IW Component}} \nonumber \\
&\quad - 
\underbrace{
\frac{I(A_i=a) - \widehat{e}_a(X_i)}{\widehat{e}_a(X_i)}
\left[
\bigl( \widehat{\mu}_a(t_0 \mid X_i) - \widehat{\mu}_a(t_0+\theta \mid X_i) \bigr)
- \tau \widehat{\mu}_a(t_0 \mid X_i)
\right]
}_{\text{Augmentation Term}}
\Bigg\} \approx 0.
\label{eq:dr_estimating_eq}
\end{align}
The estimator $\widehat{q}^{DR}_a(\tau; t_0)$ is obtained by solving equation \eqref{eq:dr_estimating_eq} with respect to $\theta$. We now state the key property of this estimator:
\begin{theorem}[Double Robustness] 
\label{thm:dr}
The estimator $\widehat{q}^{DR}_a(\tau; t_0)$ is a consistent estimator for the true quantile $q_a(\tau; t_0)$ if either the propensity score model $\widehat{e}_a$ or the outcome regression model $\widehat{\mu}_a$ is correctly specified, provided that the censoring mechanism $\widehat{G}_a$ is consistently estimated. 
\end{theorem}
The detailed proof of Theorem~\ref{thm:dr} is provided in the Supplementary Material.

The double robustness in Theorem~\ref{thm:dr} concerns the trade-off between the treatment assignment model $e_a$ and the outcome regression model $\mu_a$: the estimator remains consistent if either model is correctly specified, provided that the censoring mechanism $G_a(t\mid X)$ is consistently estimated. In contrast, misspecification of the censoring model generally induces bias because both the IW and augmentation components rely on IPCW to correct for informative loss to follow-up. We therefore recommend flexible semiparametric approaches for $G_a$, such as treatment-stratified Cox models or random survival forest (RSF) models \citep{ishwaran2019fast}.

\subsection{Inference}
Both the IW and DR estimators are solutions to non-smooth Z-estimation problems involving estimated nuisance parameters. While general theory for non-smooth estimating equations implies that the estimators are asymptotically normal \citep{pakes1989simulation}, deriving closed-form variance estimators is challenging.

Standard M-estimation theory dictates that the asymptotic variance of a quantile estimator depends on the sparsity function---the reciprocal of the probability density function of the residual lifetime at the true quantile \citep{koenker2005quantile}. Direct estimation of this quantity is computationally unstable and requires arbitrary smoothing parameter selection (e.g., bandwidths). Furthermore, valid inference must account for the uncertainty propagated by the estimation of high-dimensional nuisance parameters ($\widehat{e}_a, \widehat{G}_a$) \citep{parke1986pseudo}. To avoid the complexities of density estimation and to explicitly capture the uncertainty from nuisance parameters and between-arm correlation, we employ the nonparametric bootstrap \citep{tibshirani1993introduction}. The procedure is as follows:
\begin{enumerate}
    \item Resample $n$ individuals with replacement from the observed data to create a bootstrap sample $O^{(b)}$.
    \item Re-estimate the nuisance parameters $\widehat{\eta}^{(b)}$ (propensity scores, censoring weights, and outcome means) within the bootstrap sample.
    \item Solve for the quantiles $\widehat{q}_1^{(b)}$ and $\widehat{q}_0^{(b)}$ using the re-estimated weights.
    \item Compute the bootstrap difference: $\widehat{\delta}^{(b)} = \widehat{q}_1^{(b)} - \widehat{q}_0^{(b)}$.
\end{enumerate}
Repeating this process $B$ times (e.g., $B=500$), we estimate the standard error of the treatment effect, $\text{SE}(\widehat{\delta})$, as the sample standard deviation of the $B$ bootstrap differences. The $(1-\alpha)\%$ Wald-type confidence interval is given by $\widehat{\delta} \pm z_{1-\alpha/2} \times \text{SE}(\widehat{\delta})$, where $z_{1-\alpha/2}$ is the $(1-\alpha/2)$-quantile of the standard normal distribution, or alternatively, by the $(\alpha/2)$ and $(1-\alpha/2)$ quantiles of the bootstrap distribution.

\section{Supplementary Analysis: Decomposing Prognostic and Compositional Effects}
\label{sec:supplement}

As established in Section~\ref{subsec:setup}, the OSQC $\delta(\tau; t_0)$ provides a  pragmatic prognosis for the population of observed survivors. However, interpreting the mechanism behind this effect is complicated by the fact that survival to landmark time $t_0$ is itself a post-treatment selection. Because treatment influences who survives to $t_0$, the OSQC aggregates two distinct causal pathways:
\begin{enumerate}
    \item \textbf{Causal Efficacy within Strata:} The treatment's ability to extend residual life for specific individuals (e.g., those who would have survived regardless of treatment).
    \item \textbf{Compositional Selection:} The treatment's ability to alter the frailty profile of the survivor pool by keeping high-risk individuals alive who otherwise would have died.
\end{enumerate}

A naive comparison assumes that the subpopulation surviving under treatment ($A=1$) is comparable to the subpopulation surviving under control ($A=0$). However, if the treatment is effective at preventing early mortality, these two groups will differ in their latent composition. To understand the extent to which observed survival patterns are driven by compositional shifts versus individual-level benefits, we conduct a supplementary analysis aiming to recover the \emph{Principal Survivor Quantile Contrast} (PSQC). This estimand is analogous to the Survivor Average Causal Effect \citep{zhang2003estimation,rubin2006causal} but targets the quantile scale within the latent stratum of always-survivors.

\subsection{Principal Stratification and Reweighting}
We frame this problem using Principal Stratification \citep{frangakis2002principal, martinussen2020subtleties}. Let $S_a = I(T_a > t_0)$ be the potential indicator of survival to the landmark time under treatment $a$. The population can be partitioned into four latent principal strata based on the joint values of $(S_0, S_1)$:
\begin{itemize}
    \item \textbf{Always Survivors} $(S_0=1, S_1=1)$: Individuals robust enough to survive to $t_0$ regardless of treatment.
    \item \textbf{Protected} $(S_0=0, S_1=1)$: Individuals who survive to $t_0$ only if treated but would die before $t_0$ if untreated.
    \item \textbf{Harmed} $(S_0=1, S_1=0)$: Individuals who survive under control but die under treatment.
    \item \textbf{Never Survivors} $(S_0=0, S_1=0)$: Individuals who die before $t_0$ under either condition.
\end{itemize}

Within this framework, we define the PSQC as a causal estimand targeting the distribution of residual lifetime among the latent subpopulation of individuals who would survive to the landmark time under either treatment. The PSQC is defined as
\[
\delta^{PS}(\tau; t_0)
=
q^{PS}_1(\tau; t_0) - q^{PS}_0(\tau; t_0),
\]
where
\[
q^{PS}_a(\tau; t_0)
=
\inf\left\{ r \ge 0 :
P\!\left(T_a \le t_0 + r \mid S_0 = 1,\, S_1 = 1\right) \ge \tau
\right\}.
\]
This estimand characterizes the $\tau$-th quantile of the residual lifetime within the latent subpopulation of individuals who would survive to the landmark time under either treatment. In contrast to the OSQC, which characterizes the residual-life distribution among the observed survivors at time $t_0$ via \eqref{eq:FR_def}-\eqref{eq:delta_def}, the PSQC isolates the causal effect within a latent subpopulation that would survive to $t_0$ under either treatment, thereby removing compositional differences induced by differential survival.

Identification of this estimand requires additional assumptions beyond those needed for the OSQC, which we now state. In many clinical settings, it is plausible to assume \emph{Monotonicity}: treatment may extend survival but never shortens it ($S_1 \ge S_0$) \citep{zhang2003estimation}. This assumption implies that the ``Harmed'' stratum is empty. Consequently, the observed survivors in the control arm are purely ``Always Survivors,'' whereas survivors in the treatment arm are a mix of ``Always Survivors'' and ``Protected.'' A risk of bias arises if the ``Protected'' group has a different residual lifetime distribution than the ``Always Survivors.'' For example, if the potential residual lifetime $R_1(t_0)$ for the ``Protected'' group is shorter than that of the ``Always Survivors,'' the inclusion of these frailer individuals dilutes the aggregate benefit observed in the treatment arm.

To address this, we rely on the additional assumption of \emph{Principal Ignorability} conditional on $X$. This implies that, within levels of covariates $X$, the potential residual lifetime $R_a$ is independent of the latent stratum membership \citep{ding2017principal, feller2017principal}. Practically, this suggests that if we can identify ``Always Survivors'' in the $A=1$ arm based on their covariates, they are comparable to the survivors in the $A=0$ arm. Under these preconditions, we can reweight the treated survivors to match the covariate distribution of the control survivors (the Always Survivors). The probability of belonging to the Always Survivor stratum given survival under treatment is:
\[
P(\text{Always Survivor} \mid S_1=1, X) = \frac{P(S_0=1 \mid X)}{P(S_1=1 \mid X)}.
\]
Therefore, following the identification strategy of \citet{tchetgen2014identification}, the additional weights are defined as:
\begin{equation}
w_{sel}(A, X) = 
\begin{cases} 
1 & \text{if } A=0, \\
\dfrac{P(T_0 > t_0 \mid X)}{P(T_1 > t_0 \mid X)} & \text{if } A=1.
\end{cases}
\label{eq:selection_weights}
\end{equation}
This ratio, effectively a \emph{principal score} weighting \citep{ding2017principal}, down-weights the ``Protected'' survivors in the treatment arm to recover the latent distribution of the ``Always Survivors.'' Let $\theta_{PSQC}$ be the true $\tau$-th quantile of the residual lifetime distribution among the latent subpopulation of ``Always Survivors'' (individuals who would survive to $t_0$ under both $A=0$ and $A=1$). In this supplementary analysis, the PSQC can be identified based on the following:
\begin{theorem}[Identification of PSQC Quantile]
\label{thm:PSQC} 
Assume the Assumptions \ref{ass:consistency}-\ref{ass:censoring} hold. Further, we assume the following conditions:
\begin{assumption}[Monotonicity]
\label{ass:monotonicity}
$S_1\ge S_0$ almost surely.
\end{assumption}

\begin{assumption}[Principal Ignorability]
\label{ass:principal}
For all measurable function $h$,
\[
E\!\left[h(R_1(t_0))\mid S_0=1,S_1=1,X\right]
=
E\!\left[h(R_1(t_0))\mid S_1=1,X\right].
\]
\end{assumption}
Then, the supplementary estimator $q_{a}^{PS}$ is identified as the unique solution $\theta$ to the population moment condition
\begin{align}
E\!\left[
w_{sel}(a,X)\,
\frac{I(A=a)}{e_a(X)}
\left\{
\frac{ I(t_0<Y\le t_0+\theta,\ \Delta=1) }{ G_a(Y\mid X) }
-
\tau\,
\frac{ I(Y>t_0) }{ G_a(t_0\mid X) }
\right\}
\right]=0.
\label{eq:PSQC_moment_IW}
\end{align}
\end{theorem}
The detailed proof of Theorem~\ref{thm:PSQC} is provided in the Supplementary Material as well.

\begin{remark*}[Assumption Burden and the Role of Supplementary Analysis]
Admittedly, relying on two untestable Assumptions \ref{ass:monotonicity}-\ref{ass:principal} imposes limitations. In particular, the Assumption \ref{ass:principal} not only invokes cross-world conditioning on potential survival indicators, but also places a cumulative burden on the baseline covariates $X$; specifically, the same covariate set is required for Assumption \ref{ass:exchangeability}, \ref{ass:censoring}, and \ref{ass:principal}. Requiring a single covariate set to simultaneously satisfy these roles constitutes a strong and generally untestable restriction.

Given these constraints, we employ the Principal Stratification framework strictly as a supplementary analysis rather than as a primary identification strategy. We maintain that the OSQC remains the principal quantity of clinical interest because it is defined on an observable population: neither patients nor clinicians can observe membership in latent principal strata, but they can condition on survival status at the landmark time. By targeting the residual-life distribution among observed survivors, the OSQC formalizes the standard convention underlying descriptive landmark analyses widely used in the medical literature. The supplementary analysis therefore serves not to replace this pragmatic estimand, but rather to decompose the observed prognostic contrast, offering insight into the relative contributions of causal efficacy within individuals versus compositional changes in the survivor population when such assumptions are deemed plausible.
\end{remark*}

\begin{remark*}[Why the PSQC Quantile is Identified via IW]
The identification argument in Theorem~\ref{thm:PSQC} relies on a principal-score reweighting that operates within the observed survivor set to recover the covariate distribution of the Always-Survivor stratum. For this reason, the PSQC quantile is naturally characterized through an IW estimating equation, which involves only observed survivors and standard inverse probability weights.

Although DR estimators for principal strata effects have been developed in other contexts \citep{tchetgen2014identification, isenberg2025weighting}, they typically require complex additional modeling of stratum membership probabilities. In our setting, incorporating the principal-score weight $w_{sel}(A,X)$ directly into the DR augmentation term is not straightforward, because the augmentation term is constructed using the full treatment assignment mechanism rather than survivor-restricted moments. Without additional modeling assumptions, multiplying the augmentation component by $w_{sel}(A,X)$ does not generally preserve its mean-zero property because the augmentation term is defined with respect to the full treatment assignment mechanism rather than survivor-restricted moments. We therefore focus on the IW formulation, which aligns directly with the identification strategy for the PSQC quantile and offers a transparent mechanism for supplementary checking.
\end{remark*}

\subsection{Implementation}

We implement this supplementary analysis by incorporating the estimated selection weights $\widehat{w}_{sel}(A_i, X_i)$ into the estimating equations. The procedure consists of three steps. First, we estimate the conditional survival probabilities $\widehat{\pi}_a(X) = \widehat{P}(T > t_0 \mid A=a, X)$ using the same survival models employed for the outcome regression in Section~\ref{sec:estimation}, which implies that the estimator relies on correct specification of the outcome model. Second, we compute the weights $\widehat{w}_{sel, i}$ for all surviving subjects. For the control arm ($A=0$), the weights are fixed at 1. For the treatment arm ($A=1$), the weights are defined by $\widehat{\pi}_0(X_i)/\widehat{\pi}_1(X_i)$. Finally, we solve the reweighted estimating equation leveraging Theorem~\ref{thm:PSQC}:
\begin{equation}\label{eq:sup_estimating_eq}
    \sum_{i=1}^n \widehat{w}_{sel, i} \cdot U^{IW}_i(\theta; \widehat{\eta}) \approx 0.
\end{equation}
The estimator $\widehat{q}^{PS}_a(\tau; t_0)$ is obtained by solving equation \eqref{eq:sup_estimating_eq} for $\theta$.

Comparing the standard DR estimate ($\widehat{\delta}^{DR}$) with the supplementary-weighted estimate ($\widehat{\delta}^{PS}$) allows us to decompose the observed prognostic benefit. Specifically, if we observe that the OSQC is smaller than the PSQC (i.e., $\widehat{\delta}^{DR} < \widehat{\delta}^{PS}$), it indicates that the treatment successfully preserves the lives of frailer individuals (the ``Protected'' stratum) who otherwise would have died. Since these frail survivors likely have shorter residual lifespans, their inclusion lowers the aggregate quantile of the overall survivor population. Conversely, if the estimates are similar, it implies that the treatment effect is uniform across strata and not heavily influenced by compositional heterogeneity in the survivor pool.

\section{Simulation Study}
\label{sec:simulation}

We conducted a comprehensive simulation study (1) to verify the double robustness property, (2) to assess the finite-sample performance of the proposed estimators, and (3) to demonstrate the practical usage of the supplementary analysis tool. The data generation process (DGP) was designed to induce both confounding by baseline covariates and informative censoring, mimicking complex observational data structures. Details in the DGP is provided in Supplementary Material.

\subsection{Simulation Setup}

We conducted $1000$ Monte Carlo simulations with sample sizes of $N=500$ and $N=2000$ to evaluate asymptotic convergence. Within each simulation, we employed $1000$ bootstrap resamples to estimate standard errors and construct confidence intervals. Three baseline covariates $X_i = (X_{i1}, X_{i2}, X_{i3})^\top$ were drawn from a multivariate normal distribution with mean zero and pairwise correlation $\rho=0.2$.

Treatment assignment $A_i$ was generated from a logistic model including a non-linear quadratic term to challenge the misspecified models:
\[
\text{logit}~P(A_i=1 \mid X_i) = -0.5 + 0.5 X_{i1} + 0.5 X_{i2} - 0.2 X_{i3} + X_{i1}^2.
\]
Event times $T_i$ were generated based on a Weibull distribution with shape parameter $\nu=1.5$ and a scale parameter $\lambda_T(X_i, A_i)$ defined by:
\[
\log \lambda_T(X_i, A_i) = -1 + \beta_{\text{t}} A_i + 0.5 X_{i1} + 0.2 X_{i2} + X_{i1}^2.
\]
Censoring times $C_i$ were generated from an exponential distribution with rate $\lambda_C(X_i, A_i)$ defined by:
\[
\log \lambda_C(X_i, A_i) = -0.5 - 1.0 A_i + 0.1 X_{i3} + 0.1 (X_{i1} X_{i3}).
\]
This mechanism induced strong differential censoring, with the treatment and control groups experiencing variable censoring rates depending on covariates. The parameter $\beta_{\text{t}}\in\{-0.5,0\}$ controls the treatment effect on the true event time $T_{i}$, which strongly affects both OSQC and PSQC. While $\beta_{\text{t}}=-0.5$ assumes a standard scenario with a positive treatment effect, we also consider $\beta_{\text{t}}=0$ as a benchmark case of uniform treatment effect across strata without selection bias to validate the supplementary analysis tool. Further details on the DGP, including the mechanism ensuring key Assumptions \ref{ass:consistency}-\ref{ass:censoring} and \ref{ass:monotonicity}-\ref{ass:principal}, are provided in the Supplementary Material.

We evaluated four estimators for the residual lifetime quantiles (e.g., $\tau \in \{0.3, 0.5\}$) at landmark time $t_0=0.5$: (1) Naive KM, (2) IW, (3) DR, and (4) PS. To verify double robustness, we considered four scenarios regarding the model specification given correctly specified censoring model $G_{a}(\cdot)$:
\begin{itemize}
    \item \textbf{CC (Both Correct):} Both propensity and outcome models include the necessary quadratic term $X_{i1}^2$.
    \item \textbf{CI (Outcome Misspecified):} The outcome model omits $X_{i1}^2$, but the propensity model is correct.
    \item \textbf{IC (PS Misspecified):} The propensity model omits $X_{i1}^2$, but the outcome model is correct.
    \item \textbf{II (Both Misspecified):} Both models are incorrect.
\end{itemize}

\subsection{Simulation Results}

\begin{table}[ht]
\centering
\caption{Simulation results for landmark time $t_0=0.5$, reporting Bias, Standard Error (SE), and Wald-type 95\% Coverage Probability (CP). Bias is calculated relative to the true OSQC for KM, IW, DR and true PSQC for PS, where true values were calculated using $10^7$ Monte Carlo simulations. Note that results of the KM estimator are the same across all scenarios. \textbf{Abbreviations:} CC, both propensity and outcome models correct; CI, correct propensity but misspecified outcome; IC, correct outcome but misspecified propensity; II, both misspecified.}\label{tab:sim_results_t0.5}
\begin{tabular}{ll | ccc  ccc | ccc ccc}
\hline\hline
\multicolumn{2}{l|}{\multirow{2}{*}{\textbf{\shortstack{Landmark \\ $t_{0}=0.5$}}}} & \multicolumn{6}{c|}{$\tau=0.3$} & \multicolumn{6}{c}{$\tau=0.5$} \\
\multicolumn{2}{l|}{} & \multicolumn{3}{c}{\textbf{$N=500$}} & \multicolumn{3}{c|}{$N=2000$} & \multicolumn{3}{c}{$N=500$} & \multicolumn{3}{c}{$N=2000$} \\  
\cmidrule(lr){3-5} \cmidrule(lr){6-8} \cmidrule(lr){9-11} \cmidrule(lr){12-14}
\multicolumn{2}{l|}{\textbf{\shortstack{Scenario \\ $\&$Method}}} & \textbf{Bias} & \textbf{SE} & \textbf{CP} & \textbf{Bias} & \textbf{SE} & \textbf{CP} & \textbf{Bias} & \textbf{SE} & \textbf{CP} & \textbf{Bias} & \textbf{SE} & \textbf{CP} \\ \hline
\multicolumn{2}{c}{\textit{ $\beta_{\text{t}} = 0$}} & \multicolumn{6}{c}{\textit{OSQC = PSQC = 0}} & \multicolumn{6}{c}{\textit{OSQC = PSQC = 0}} \\
\hline
\multirow{4}{*}{CC}
 & KM  & -0.39 & 0.30 & 0.81 & -0.37 & 0.14 & 0.31 & -0.28 & 0.19 & 0.74 & -0.28 & 0.09 & 0.15 \\
 & IW & -0.01 & 0.15 & 0.95 & -0.01 & 0.07 & 0.95 & -0.00 & 0.23 & 0.94 & -0.01 & 0.10 & 0.95 \\
 & DR  & -0.01 & 0.15 & 0.95 & -0.01 & 0.07 & 0.95 & -0.01 & 0.22 & 0.95 & -0.01 & 0.10 & 0.95 \\
 & PS & -0.01 & 0.15 & 0.95 & -0.01 & 0.07 & 0.95 & -0.00 & 0.22 & 0.94 & -0.01 & 0.10 & 0.95 \\ \hline
\multirow{3}{*}{CI}
 & IW & -0.01 & 0.15 & 0.95 & -0.01 & 0.07 & 0.95 & -0.00 & 0.23 & 0.94 & -0.01 & 0.10 & 0.95 \\
 & DR  & -0.01 & 0.18 & 0.96 & -0.01 & 0.08 & 0.96 & -0.00 & 0.25 & 0.94 & -0.01 & 0.14 & 0.95 \\
 & PS & -0.03 & 0.15 & 0.94 & -0.02 & 0.07 & 0.94 & -0.02 & 0.22 & 0.95 & -0.03 & 0.10 & 0.95 \\ \hline
\multirow{3}{*}{IC}
 & IW & -0.18 & 0.14 & 0.77 & -0.17 & 0.07 & 0.29 & -0.24 & 0.21 & 0.82 & -0.24 & 0.10 & 0.35 \\
 & DR  & -0.01 & 0.13 & 0.95 & -0.01 & 0.06 & 0.95 & -0.01 & 0.21 & 0.95 & -0.01 & 0.10 & 0.96 \\
 & PS & -0.18 & 0.14 & 0.77 & -0.17 & 0.07 & 0.29 & -0.24 & 0.21 & 0.82 & -0.24 & 0.10 & 0.35 \\ \hline
\multirow{3}{*}{II}
 & IW & -0.18 & 0.14 & 0.77 & -0.17 & 0.07 & 0.29 & -0.24 & 0.21 & 0.82 & -0.24 & 0.10 & 0.35 \\
 & DR  & -0.19 & 0.14 & 0.75 & -0.18 & 0.07 & 0.24 & -0.27 & 0.21 & 0.78 & -0.26 & 0.10 & 0.25 \\
 & PS & -0.19 & 0.14 & 0.73 & -0.19 & 0.07 & 0.22 & -0.26 & 0.21 & 0.80 & -0.26 & 0.10 & 0.28 \\ \hline
 \multicolumn{2}{c}{\textit{ $\beta_{\text{t}} = -0.5$}} & \multicolumn{6}{c}{\textit{OSQC $\approx$ 0.22, \quad PSQC $\approx$ 0.27}} & \multicolumn{6}{c}{\textit{OSQC $\approx$ 0.39, \quad PSQC $\approx$ 0.45}} \\
 \hline
\multirow{4}{*}{CC}
 & KM  & -0.08 & 0.33 & 0.96 & -0.06 & 0.15 & 0.94 & -0.36 & 0.21 & 0.66 & -0.36 & 0.10 & 0.06 \\
 & IW & -0.01 & 0.16 & 0.94 & -0.01 & 0.08 & 0.95 & 0.00 & 0.25 & 0.95 & -0.00 & 0.12 & 0.95 \\
 & DR  & -0.01 & 0.16 & 0.95 & -0.01 & 0.08 & 0.96 & 0.00 & 0.25 & 0.96 & -0.00 & 0.11 & 0.95 \\
 & PS & -0.01 & 0.17 & 0.95 & -0.01 & 0.08 & 0.96 & 0.00 & 0.26 & 0.96 & -0.00 & 0.12 & 0.96 \\ \hline
\multirow{3}{*}{CI}
 & IW & -0.01 & 0.16 & 0.94 & -0.01 & 0.08 & 0.95 & 0.00 & 0.25 & 0.95 & -0.00 & 0.12 & 0.95 \\
 & DR  & -0.01 & 0.18 & 0.96 & -0.01 & 0.09 & 0.96 & 0.00 & 0.28 & 0.95 & -0.01 & 0.15 & 0.95 \\
 & PS & -0.08 & 0.16 & 0.92 & -0.08 & 0.08 & 0.85 & -0.08 & 0.25 & 0.94 & -0.09 & 0.12 & 0.88 \\ \hline
\multirow{3}{*}{IC}
 & IW & -0.23 & 0.15 & 0.70 & -0.22 & 0.07 & 0.17 & -0.31 & 0.23 & 0.75 & -0.30 & 0.11 & 0.22 \\
 & DR  & -0.01 & 0.15 & 0.95 & -0.01 & 0.07 & 0.96 & -0.00 & 0.23 & 0.95 & -0.00 & 0.11 & 0.96 \\
 & PS & -0.21 & 0.15 & 0.73 & -0.21 & 0.08 & 0.24 & -0.29 & 0.24 & 0.79 & -0.28 & 0.11 & 0.31 \\ \hline
\multirow{3}{*}{II}
 & IW & -0.23 & 0.15 & 0.70 & -0.22 & 0.07 & 0.17 & -0.31 & 0.23 & 0.75 & -0.30 & 0.11 & 0.22 \\
 & DR  & -0.24 & 0.15 & 0.67 & -0.24 & 0.07 & 0.11 & -0.35 & 0.23 & 0.70 & -0.34 & 0.11 & 0.12 \\
 & PS & -0.29 & 0.15 & 0.53 & -0.29 & 0.07 & 0.03 & -0.40 & 0.23 & 0.64 & -0.39 & 0.11 & 0.06 \\ \hline\hline
\end{tabular}
\end{table}

The simulation results for the primary landmark time $t_0=0.5$ are summarized in Table~\ref{tab:sim_results_t0.5}; additional results for an alternative landmark time ($t_0=0.3$) showing consistent patterns are provided in the Supplementary Material. First, our findings unequivocally validate the double robustness of the proposed DR estimator. Under the current framework, the KM exhibits substantial bias across all scenarios due to its failure to adjust for confounding (e.g., Bias $\approx -0.37$ for $N=2000$ at $\tau=0.3$). While the IW estimator performs effectively when the propensity score is correctly specified, it incurs significant bias in the \textbf{IC} and \textbf{II} scenarios where the propensity model is misspecified (e.g., Bias $\approx -0.17$ for $N=2000$ at $\tau=0.3$). In contrast, the DR estimator maintains unbiasedness (Bias $\leq \pm 0.01$) across the three robust scenarios (\textbf{CC}, \textbf{CI}, and \textbf{IC}) by successfully leveraging the correct specification of either the propensity or the outcome model. Notably, as the sample size increases from $N=500$ to $N=2000$, the bias of the DR estimator remains negligible while its precision improves, underscoring its asymptotic consistency. As expected, all estimators exhibit bias in the \textbf{II} scenario, where both models are misspecified. It should be emphasized that these results are predicated on the correct specification of the censoring model, which remains a fundamental prerequisite for unbiased estimation in this setting.

Regarding statistical inference, the proposed estimators (IW, DR, and PS) demonstrate excellent finite-sample properties within their respective valid model specifications—specifically, IW under \textbf{CC} and \textbf{CI}; DR under \textbf{CC}, \textbf{CI}, and \textbf{IC}; and PS under \textbf{CC}. Consistent with theoretical expectations, the estimators exhibit a clear $\sqrt{n}$-convergence rate, with standard errors decreasing by approximately 50\% as the sample size increases fourfold from $N=500$ to $N=2000$ (e.g., from $0.15$ to $0.07$ for the DR estimator at $\tau=0.3$ under \textbf{CC}). Furthermore, the empirical coverage probabilities (CP) for these estimators consistently approach the nominal 95\% level as $N$ grows. Most significantly, the DR estimator maintains valid coverage ($\text{CP} \approx 0.95$) across all robust scenarios, suggesting that the proposed bootstrap technique provides a reliable basis for inference even under partial model misspecification.

Finally, the comparison between the OSQC and the PSQC underscores a fundamental structural divergence between observed and latent estimands, rather than mere estimation bias. In the null case (Panel A: $\beta_{\text{t}}=0$), where selection bias is absent, the true OSQC and PSQC are equivalent (both equal to zero); consequently, both the DR and PS estimators converge to the same null value. However, in the presence of a protective treatment effect (Panel B: $\beta_{\text{t}}=-0.5$), the induction of a ``Protected'' stratum causes the true parameters to deviate. The PS estimator successfully identifies the latent PSQC parameter only when all nuisance functions—propensity, outcome, and censoring models—are correctly specified (scenario \textbf{CC}). 

The vulnerability of the PS estimator is particularly evident under model misspecification; for instance, in the \textbf{IC} scenario at $\tau=0.5$, the bias remains substantial (Bias $\approx -0.28$ for $N=2000$) even as precision improves markedly with sample size (SE decreasing from $0.24$ at $N=500$ to $0.11$ at $N=2000$). Furthermore, the PS estimator relies on the stringent and untestable assumptions of monotonicity and principal ignorability. Given the inherent difficulty in satisfying these causal assumptions alongside complex model specifications in practice, obtaining reliable PSQC estimates remains a significant challenge. These results reinforce the utility of the proposed DR estimator as a more practical and robust framework across varied scenarios, whereas the PS estimator may be more appropriately positioned as a specialized tool for supplementary analysis.

\section{Application}\label{sec:application}
\subsection{Application to the SUPPORT Right Heart Catheterization Study}\label{sec:app1}

We illustrate the proposed landmark residual-life quantile estimator using data from the SUPPORT (Study to Understand Prognoses and Preferences for Outcomes and Risks of Treatments) right heart catheterization (RHC) study originally analyzed by \citet{connors1996effectiveness}. This dataset serves as a canonical benchmark in the causal inference literature for evaluating methods under high-dimensional confounding \citep{hirano2001estimation, tan2006distributional}. The study followed critically ill adults admitted to intensive care units (ICU) in the United States and compared outcomes between patients who received RHC within 24 hours of admission and those who did not. Detailed clinical background and study design are provided in the original publication \citet{connors1996effectiveness}.

Our estimand, OSQC, targets the treatment effect of RHC on conditional residual lifetime, defined as the $\tau$-quantile of remaining survival time beyond a prespecified landmark $t_0$, among patients who survive to $t_0$ under their observed treatment. This estimand addresses how RHC affects subsequent prognosis among patients who stabilize beyond the early acute phase. Motivated by the highly front-loaded mortality in the SUPPORT cohort, we consider landmark times $t_0 = 3, 7,$ and $14$ days. The landmark $t_0=3$ days captures an early stabilization phase immediately following a sharp mortality decline between days 2 and 3; $t_0=7$ days conditions on survival beyond the acute ICU phase; and $t_0=14$ days focuses on a more selective group of longer-term survivors, shifting the estimand toward prognosis in a recovering or clinically complicated cohort.

We focus on the lower-tail quantile $\tau=0.3$ rather than the median because the median residual lifetime is not identifiable within the 180-day follow-up horizon: approximately 60\% of control patients remain alive at administrative censoring. In contrast, the $30^{\text{th}}$ percentile is estimable across all landmark times and captures clinically meaningful variation in poor-prognosis outcomes relevant to critically ill populations.

Nuisance functions were specified to mirror the modeling strategy of \citet{connors1996effectiveness}, adjusting for a rich set of baseline covariates including age, sex, APACHE III score, mean blood pressure, and diagnostic categories. The propensity score for RHC receipt was estimated using logistic regression, while censoring was modeled using a Cox proportional hazards model conditional on treatment and covariates. The outcome regression component was specified as a stratified Cox model with treatment-specific baseline hazards and shared covariate effects, avoiding proportional hazards assumptions for the treatment effect.

\begin{table}[t]
\centering
\caption{Estimated treatment effect of RHC on the $\tau=0.3$ residual-life quantile contrast
$\widehat{\delta}(\tau;t_0)=\widehat q_{1}(\tau;t_0)-\widehat q_{0}(\tau;t_0)$ (days).
Standard errors are estimated using $B=1000$ nonparametric bootstrap resamples.
Wald-type 95\% confidence intervals are reported.
Negative values indicate shorter residual life under RHC among patients surviving to landmark time $t_0$ (days).}
\label{tab:rhc_app_wald}
\begin{tabular}{l c c c c}
\hline\hline
$t_0$ (days) & Estimator & $\widehat{\delta}(\tau;t_0)$ & Bootstrap SE & 95\% Wald CI \\
\hline
3  & DR  & $-11.0$ & $4.75$ & $[-20.31,\,-1.69]$ \\
   & IW & $-10.0$ & $4.46$ & $[-18.74,\,-1.26]$ \\
\hline
7  & DR  & $-25.0$ & $9.86$ & $[-44.33,\,-5.67]$ \\
   & IW & $-22.0$ & $9.28$ & $[-40.19,\,-3.81]$ \\
\hline
14 & DR  & $-42.0$ & $23.54$ & $[-88.15,\,4.15]$ \\
   & IW & $-41.0$ & $22.65$ & $[-85.38,\,3.38]$ \\
\hline\hline
\end{tabular}
\end{table}

Table~\ref{tab:rhc_app_wald} reports treatment effect estimates using the DR estimator as the primary analysis, with the IW estimator provided for comparison. Inference was conducted using a nonparametric bootstrap with 1,000 resamples. At $t_0=3$ days, the DR estimate of the $30^{\text{th}}$ percentile residual-life contrast was $-11.0$ days (95\% CI: $[-20.31,\,-1.69]$), indicating shorter lower-tail residual survival among RHC recipients conditional on surviving three days. At $t_0=7$ days, the estimated contrast widened to $-25.0$ days (95\% CI: $[-44.33,\,-5.67]$). At $t_0=14$ days, although the point estimate suggested a large deficit ($-42.0$ days), uncertainty increased substantially and the confidence interval included zero, reflecting the reduced effective sample size among longer-term survivors.

These findings are qualitatively consistent with prior analyses of the SUPPORT cohort, which reported worse outcomes among patients receiving RHC. Our results refine this understanding by quantifying the deficit in residual life specifically among survivors of the acute phase. We do not report the PS estimator for this application because the monotonicity assumption (that RHC never harms survival) can be implausible in this setting, as invasive procedures may carry heterogeneous risks that shorten survival for certain frail subgroups.

\subsection{Application to the NSABP B-14 Tamoxifen Trial}
\label{sec:app2}

We illustrate the proposed landmark residual-life quantile estimators using data from the National Surgical Adjuvant Breast and Bowel Project (NSABP) B-14 randomized clinical trial comparing adjuvant tamoxifen versus placebo in estrogen receptor--positive, node-negative breast cancer, originally reported in the clinical literature \citep{Fisher2004Lancet} and subsequently analyzed in the statistical literature \citep{Jeong2006JRSSA}. The trial enrolled women following definitive surgery and compared subsequent disease-related outcomes between the tamoxifen and placebo arms under longitudinal follow-up. In the present analysis, the dataset contains five variables: treatment assignment, age in years, tumor size in centimeters, follow-up time since surgery in years, and an event indicator. Although treatment was randomized, some subjects were excluded from the analytic cohort due to missingness; accordingly, we adopt a working propensity-score model to demonstrate the proposed estimation pipeline in a setting where randomization may not be perfectly preserved in the analyzed sample.

Again our primary estimand is the OSQC, evaluated at landmark times $t_0\in\{1,2,3,4,5\}$ and the lower-tail quantile $\tau=0.3$ to emphasize clinically meaningful differences in poor-prognosis outcomes and to maintain stable estimation across later landmark times, where approximately half of the patients were censored.

Nuisance functions were specified using the baseline covariates (age and tumor size). The propensity score for tamoxifen receipt was estimated using the logistic regression. To account for informative loss to follow-up, we modeled censoring conditional on treatment and covariates using two approaches: (i) a Cox proportional hazards model for the censoring time, and (ii) a RSF model as a robustness check, which we did not utilize in the RHC application due to a larger data size with over 60 covariates. The outcome regression component was specified as a stratified Cox model for the event time with treatment-specific baseline hazards and shared covariate effects, which avoids imposing a proportional hazards restriction on the treatment effect. In addition to OSQC, we also report an illustrative PSQC using the supplementary estimator introduced in Section~\ref{sec:supplement}.

\begin{table}[t]
\centering
\caption{Estimated treatment effect of tamoxifen on the $\tau=0.3$ residual-life
quantile contrast $\widehat{\delta}(\tau;t_0)=\widehat q_{1}(\tau;t_0)-\widehat q_{0}(\tau;t_0)$ (years) across landmark times $t_0$.
For each estimator, results are shown under two censoring models (Cox and RSF), with RSF included as a robustness check.
Standard errors are estimated using $B=1000$ nonparametric bootstrap resamples,
and Wald-type 95\% confidence intervals are reported.
Positive values indicate longer residual life under tamoxifen among patients surviving to landmark time $t_0$.}
\label{tab:b14_app_wald}
\begin{tabular}{l l c c c c c c}
\hline\hline
\;$t_0$ & Estimator &
\multicolumn{3}{c}{Cox censoring model} & \multicolumn{3}{c}{RSF censoring model} \\
\cmidrule(lr){3-5} \cmidrule(lr){6-8}
(yrs) & &
$\widehat{\delta}(\tau;t_0)$ & Bootstrap SE & 95\% Wald CI &
$\widehat{\delta}(\tau;t_0)$ & Bootstrap SE & 95\% Wald CI \\
\hline
1 & DR & $3.31$ & $0.66$ & $[2.02,\,4.59]$ & $3.47$ & $0.69$ & $[2.11,\,4.82]$ \\
  & IW & $3.31$ & $0.66$ & $[2.02,\,4.59]$ & $3.47$ & $0.69$ & $[2.11,\,4.82]$ \\
  & PS & $3.31$ & $0.65$ & $[2.03,\,4.59]$ & $3.47$ & $0.69$ & $[2.11,\,4.82]$ \\
\hline
2 & DR & $2.73$ & $0.64$ & $[1.47,\,3.99]$ & $2.72$ & $0.65$ & $[1.45,\,4.00]$ \\
  & IW & $2.73$ & $0.64$ & $[1.47,\,4.00]$ & $2.72$ & $0.65$ & $[1.45,\,4.00]$ \\
  & PS & $2.73$ & $0.64$ & $[1.47,\,4.00]$ & $2.74$ & $0.65$ & $[1.46,\,4.03]$ \\
\hline
3 & DR & $2.37$ & $0.69$ & $[1.02,\,3.73]$ & $2.49$ & $0.74$ & $[1.05,\,3.94]$ \\
  & IW & $2.37$ & $0.69$ & $[1.02,\,3.73]$ & $2.49$ & $0.74$ & $[1.05,\,3.94]$ \\
  & PS & $2.37$ & $0.69$ & $[1.02,\,3.73]$ & $2.50$ & $0.74$ & $[1.06,\,3.94]$ \\
\hline
4 & DR & $1.39$ & $0.72$ & $[-0.03,\,2.81]$ & $1.33$ & $0.72$ & $[-0.07,\,2.74]$ \\
  & IW & $1.39$ & $0.73$ & $[-0.03,\,2.81]$ & $1.33$ & $0.72$ & $[-0.07,\,2.74]$ \\
  & PS & $1.39$ & $0.72$ & $[-0.03,\,2.81]$ & $1.33$ & $0.72$ & $[-0.07,\,2.74]$ \\
\hline
5 & DR & $0.88$ & $0.58$ & $[-0.26,\,2.01]$ & $0.86$ & $0.64$ & $[-0.39,\,2.11]$ \\
  & IW & $0.88$ & $0.58$ & $[-0.26,\,2.01]$ & $0.87$ & $0.64$ & $[-0.37,\,2.12]$ \\
  & PS & $0.88$ & $0.58$ & $[-0.26,\,2.01]$ & $0.86$ & $0.63$ & $[-0.38,\,2.10]$ \\
\hline\hline
\end{tabular}
\end{table}

Table~\ref{tab:b14_app_wald} reports estimated treatment effects using the DR estimator as the primary OSQC analysis, with the IW estimator provided for comparison and the PS estimator included as an illustrative PSQC analysis. Inference was conducted using $B=1000$ nonparametric bootstrap resamples. At the earliest landmark $t_0=1$ year, the estimated $\tau=0.3$ residual-life contrast was $3.3$ years under the Cox censoring model (95\% CI: $[2.02,\,4.59]$) and $3.5$ years under the RSF model (95\% CI: $[2.11,\,4.82]$). The magnitude of the contrast declined with later landmarks, with estimates of approximately $2.4$--$2.5$ years at $t_0=3$ and $0.9$ years at $t_0=5$. For $t_0\in\{4,5\}$ years, confidence intervals included zero, reflecting increased uncertainty due to progressive selection and reduced effective sample size among later landmark survivors.

Across all landmark times, estimates were highly stable across nuisance specifications: the IW, DR, and illustrative PS estimators were nearly identical, and results obtained under Cox and RSF censoring models were closely aligned. Under the maintained monotonicity and principal ignorability assumptions, the similarity between OSQC and PSQC estimates suggests that selection effects induced by conditioning on survival to the landmark time are minimal in this dataset. Likewise, the agreement between Cox- and RSF-based analyses indicates limited sensitivity to the censoring model, suggesting little evidence of substantial model misspecification. Overall, these findings support the robustness of the proposed framework and indicate that, among individuals surviving to the landmark time, tamoxifen is associated with longer lower-tail residual lifetime, with the effect attenuating at later landmarks.

\section{Discussion}
\label{sec:discussion}

We developed a causal framework for quantiles of residual lifetime that targets clinically interpretable prognosis among individuals who survive to a prespecified landmark time $t_0$. Our primary estimand, the OSQC, contrasts residual-life quantiles between treatment arms within the survivor population at $t_0$, with explicit indexing by both the landmark time $t_0$ and the quantile level $\tau$. This formulation directly addresses a common clinical question: conditional on having reached a meaningful milestone, how does treatment affect remaining survival? By focusing on quantiles rather than marginal survival contrasts, the OSQC summarizes post-$t_0$ prognosis in a way that is robust to skewness and tailored to decision-relevant regions of the outcome distribution, such as lower-tail quantiles.

Because survival to the landmark time is itself affected by treatment, the OSQC necessarily compares dynamically selected survivor populations rather than a fixed baseline cohort. Rather than viewing this post-treatment conditioning as a limitation, we interpret the resulting contrast as describing treatment effects within the clinically relevant population of individuals who are alive and under observation at time $t_0$. At the same time, although the primary estimand conditions on survival, it can be still of interest to understand how much of the observed contrast reflects compositional changes in the survivor population. To address this question, we introduced a supplementary analysis based on principal stratification, which targets the PSQC. This analysis relies on additional, untestable Assumptions \ref{ass:monotonicity}-\ref{ass:principal} and should therefore be interpreted cautiously. We view the PSQC not as a replacement for the primary estimand, but as a complementary tool that helps disentangle the role of selective survival when such assumptions are deemed plausible. By focusing on the OSQC, we avoid relying primarily on such latent constructs, while acknowledging that the resulting estimand reflects the functional impact of intervention on the evolving survivor population.

The proposed DR estimator augments the IW estimator with a working outcome regression and is consistent when either the propensity score model or the outcome regression model is correctly specified, provided the censoring mechanism is consistently estimated. Our simulation studies supported these theoretical properties, with the DR estimator exhibiting negligible bias and near-nominal coverage across a range of scenarios in which at least one nuisance model was correctly specified. Because quantile estimating equations are inherently non-smooth and analytical variance estimators based on density estimation can be unstable, we adopted nonparametric bootstrap inference. The bootstrap performed well in finite samples and achieved close-to-nominal coverage even under misspecification of one nuisance model, representing a practical advantage over analytical variance estimators in this setting despite the increased computational cost.

Our application to the SUPPORT RHC study illustrates the practical use of residual-life quantiles in an observational setting with heavy early mortality. Focusing on $\tau=0.3$ enabled estimation within the available follow-up, and results suggested shorter lower-tail residual survival among catheterized patients at early landmarks, with increasing uncertainty at later landmarks. In contrast, the B-14 tamoxifen application demonstrates the framework in a randomized trial with longer-term follow-up, where tamoxifen was associated with longer lower-tail residual lifetime among survivors at early landmarks, with the magnitude of the contrast attenuating as the landmark time increased. Together, these applications highlight how landmark-based residual-life quantiles can flexibly characterize treatment effects on post-stabilization prognosis across diverse clinical settings.

Several important limitations and directions for future work merit discussion. First, while our supplementary analysis regarding the PSQC targets the specific issue of post-treatment selection, it does not evaluate the robustness of the proposed OSQC estimand to violations of its own identifying assumptions. Specifically, identification relies not only on conditional exchangeability and independent censoring, but also on the structural assumptions of consistency and positivity (Assumptions~\ref{ass:consistency}--\ref{ass:censoring}). A substantial body of research emphasizes that departures from these conditions can induce bias or instability. For instance, unmeasured confounding violates exchangeability \citep{robins2000sensitivity, vanderweele2017sensitivity}, while practical violations of positivity (e.g., limited covariate overlap) can lead to unstable weights and finite-sample bias \citep{petersen2012diagnosing}. Furthermore, if the treatment is not well-defined, violations of the consistency assumption may render the estimand vague \citep{cole2009consistency, vanderweele2009concerning}. Future research could therefore extend the proposed framework by developing comprehensive sensitivity analyses that directly probe departures from this full suite of identification conditions.

In addition, the proposed bootstrap-based variance estimation can impose a substantial computational burden, especially when inference is required across multiple landmark times $t_0$ and quantile levels $\tau$. To alleviate this, future work could adapt multiplier bootstrap methods, which offer an efficient alternative by approximating the limit distribution through perturbations of the estimating equations \citep{kojadinovic2012goodness, bae2025diagnostics}. Alternatively, induced smoothing techniques could be implemented to derive a closed-form variance estimator, thereby bypassing the non-smoothness of the objective function and eliminating the need for computationally intensive resampling \citep{brown2006induced, choi2024general}. Furthermore, extending Bayesian frameworks to this causal setting would provide a robust platform for exact uncertainty quantification via posterior sampling \citep{park2012bayesian}. Recent work by \citet{grossi2025bayesian} has demonstrated the utility of Bayesian principal stratification for longitudinal data subject to truncation by death, suggesting a promising avenue for integrating our quantile-based estimands into fully probabilistic models.


\section*{Disclaimer}
The views expressed in this article should not be construed to represent those of U.S. Food and Drug Administration.

\section*{Conflict of Interests}
The authors declare no conflicts of interest.

\section*{Data Availability Statement}
The simulation code and the analysis scripts for both the SUPPORT and NSABP B-14 studies are available on GitHub at \url{https://github.com/tkh5956/Causal_QRL}. The SUPPORT RHC dataset is accessible through the repository; however, the NSABP B-14 dataset is not publicly available.

    {
    \setlength{\bibsep}{0pt}
    \bibliographystyle{agsm}
    \bibliography{bibtex}
    }

\clearpage
\setcounter{section}{0}
\renewcommand*{\thetable}{\Alph{table}}
\renewcommand*{\thefigure}{\Alph{figure}}
\renewcommand*{\thesection}{\Alph{section}}

\section{Proof of Theorem \ref{thm:identification}: Identification}
\label{sec:proof_identification}

We provide the detailed derivation of the identification result presented in Theorem \ref{thm:identification} of the main text.

\subsection{Target Quantity}
Recall that the target estimand is the cumulative distribution function (CDF) of the residual lifetime $R_a(t_0) = T_a - t_0$ among those who survive to $t_0$ under treatment $a$:
\begin{equation}
F_{R_a}(r; t_0) = \frac{ P(t_0 < T_a \le t_0+r) }{ P(T_a > t_0) }.
\label{eq:S_target}
\end{equation}
We show that the numerator and denominator are uniquely identified by weighted expectations of the observed data.

\subsection{Step 1: Identification via the G-formula (Adjusting for Confounding)}
Consider the numerator probability $P(t_0 < T_a \le t_0+r)$. By the Law of Total Probability conditioning on baseline covariates $X$:
\[
P(t_0 < T_a \le t_0+r) = \int P(t_0 < T_a \le t_0+r \mid X=x) \, dP(X=x).
\]
Under \textbf{Assumption 3 (Conditional Exchangeability)}, we have $T_a \perp A \mid X$. Therefore, the counterfactual probability equates to the conditional observed probability:
\[
P(t_0 < T_a \le t_0+r \mid X=x) = P(t_0 < T \le t_0+r \mid A=a, X=x).
\]
Substituting this back, we obtain the standard G-formula representation:
\begin{equation}
P(t_0 < T_a \le t_0+r) = E_X \left[ P(t_0 < T \le t_0+r \mid A=a, X) \right].
\label{eq:S_gformula}
\end{equation}

\subsection{Step 2: Identification under Censoring (IPCW)}
The term $P(t_0 < T \le t_0+r \mid A=a, X)$ involves the true event time $T$, which is subject to right censoring $C$. We observe $Y = \min(T, C)$ and $\Delta = I(T \le C)$.

Under \textbf{Assumption 4 (Independent Censoring Given Baseline Covariates)} and the positivity condition $G_a(t\mid X)>0$ on the risk set, standard IPCW identities imply that for any measurable function $h$ and any time point $t$,
\[
E\!\left[h(T)\,I(T\le t)\mid A=a,X\right]
=
E\!\left[\frac{\Delta\,h(Y)\,I(Y\le t)}{G_a(Y\mid X)}\,\Bigg|\,A=a,X\right],
\]
where $G_a(t\mid X) = P(C\ge t\mid A=a,X,\,T\ge t)$ is the conditional probability of remaining uncensored up to time $t$ among those still under observation at $t$.

Applying this to the indicator function $h(T) = I(t_0 < T \le t_0+r)$:
\begin{align}
P(t_0 < T \le t_0+r \mid A=a, X) &= E \left[ I(t_0 < T \le t_0+r) \mid A=a, X \right] \nonumber \\
&= E \left[ \frac{I(t_0 < Y \le t_0+r, \Delta=1)}{G_a(Y \mid X)} \,\Bigg|\, A=a, X \right].
\label{eq:S_ipcw}
\end{align}

\subsection{Step 3: Combining and Reweighting}
Substituting \eqref{eq:S_ipcw} into \eqref{eq:S_gformula}, we have:
\[
P(t_0 < T_a \le t_0+r) = E_X \left[ E \left( \frac{I(t_0 < Y \le t_0+r, \Delta=1)}{G_a(Y \mid X)} \,\Bigg|\, A=a, X \right) \right].
\]
To express this nested expectation as a single expectation over the observed data distribution $P(O)$, we use the propensity score density ratio identity. For any function $g(O)$:
\[
E_X [ E(g(O) \mid A=a, X) ] = E \left[ \frac{I(A=a)}{P(A=a \mid X)} g(O) \right] = E \left[ \frac{I(A=a)}{e_a(X)} g(O) \right].
\]
Applying this, the numerator becomes:
\[
P(t_0 < T_a \le t_0+r) = E \left[ \frac{I(A=a)}{e_a(X)} \frac{I(t_0 < Y \le t_0+r, \Delta=1)}{G_a(Y \mid X)} \right].
\]
An identical derivation applies to the denominator $P(T_a > t_0)$, yielding:
\[
P(T_a > t_0) = E \left[ \frac{I(A=a)}{e_a(X)} \frac{I(Y > t_0)}{G_a(t_0 \mid X)} \right].
\]
The ratio of these two expectations yields the result in Theorem \ref{thm:identification}. \qed \\


\section{Proof of Theorem \ref{thm:dr}: Double Robustness}
\label{sec:proof_dr}

Fix a treatment arm $a\in\{0,1\}$ and a landmark time $t_0>0$.
Let the true target quantile be
\[
\theta_0 \equiv q_a(\tau;t_0),
\]
i.e., $\theta_0$ solves the population moment condition induced by the identification equation.
Throughout this proof we assume the censoring mechanism is consistently estimated, i.e.,
$\widehat G_a(\cdot\mid X) \to G_a(\cdot\mid X)$ in probability,
and we work at the population level with the limiting nuisance functions.

\paragraph{Step 1: Define the population DR estimating function in main-text notation.}
Define the observed-data ``IW score component'' for $\theta\ge 0$ by
\begin{equation}
\label{eq:dr_score_def}
H_a(O;\theta)
:=
\frac{ I(t_0<Y\le t_0+\theta,\ \Delta=1) }{ G_a(Y\mid X) }
\;-\;
\tau\,
\frac{ I(Y>t_0) }{ G_a(t_0\mid X) } .
\end{equation}
Let $e_a(X)=P(A=a\mid X)$ denote the true propensity score.
Let $\mu_a(t\mid X)=P(T>t\mid A=a,X)$ denote the true conditional survival function used in the main text.
Using $\mu_a(\cdot\mid X)$, define the outcome-regression augmentation functional
\begin{equation}
\label{eq:dr_aug_def}
m_a(X;\theta)
:=
\Bigl\{\mu_a(t_0\mid X)-\mu_a(t_0+\theta\mid X)\Bigr\}
\;-\;
\tau\,\mu_a(t_0\mid X).
\end{equation}
Note that \eqref{eq:dr_aug_def} matches exactly the bracketed term multiplying
$\{I(A=a)-\widehat e_a(X)\}/\widehat e_a(X)$ in the sample estimating equation
\eqref{eq:dr_estimating_eq} of the main manuscript, with $\widehat\mu_a$ replaced by its probability limit.

Now let $e^\dagger_a(X)$ and $\mu^\dagger_a(t\mid X)$ denote the probability limits of the fitted models
$\widehat e_a(X)$ and $\widehat\mu_a(t\mid X)$, respectively (these may be misspecified).
Define the corresponding probability-limit augmentation functional
\begin{equation}
\label{eq:dr_aug_limit_def}
m_a^\dagger(X;\theta)
:=
\Bigl\{\mu^\dagger_a(t_0\mid X)-\mu^\dagger_a(t_0+\theta\mid X)\Bigr\}
\;-\;
\tau\,\mu^\dagger_a(t_0\mid X).
\end{equation}
The population (limit) of the doubly robust estimating function is then
\begin{align}
\label{eq:U_DR_pop}
U^{DR}(\theta)
&:=
\mathbb E\!\left[
\frac{I(A=a)}{e^\dagger_a(X)}\,H_a(O;\theta)
\;-\;
\frac{I(A=a)-e^\dagger_a(X)}{e^\dagger_a(X)}\,m_a^\dagger(X;\theta)
\right].
\end{align}
Equivalently, using algebra,
\begin{align}
\label{eq:U_DR_pop_rearranged}
U^{DR}(\theta)
&=
\mathbb E\!\left[
\frac{I(A=a)}{e^\dagger_a(X)}\Bigl\{H_a(O;\theta)-m_a^\dagger(X;\theta)\Bigr\}
\;+\;
m_a^\dagger(X;\theta)
\right].
\end{align}

\paragraph{Step 2: Key unbiasedness identity under correct censoring weights.}
Under consistent estimation of $G_a(\cdot\mid X)$, the IPCW identities yield, for any $\theta\ge 0$,
\begin{align}
\label{eq:ipcw_id_1}
\mathbb E\!\left[
\frac{ I(t_0<Y\le t_0+\theta,\ \Delta=1) }{ G_a(Y\mid X) }
\Bigm| A=a, X
\right]
&=
P(t_0<T\le t_0+\theta\mid A=a,X),
\\[3pt]
\label{eq:ipcw_id_2}
\mathbb E\!\left[
\frac{ I(Y>t_0) }{ G_a(t_0\mid X) }
\Bigm| A=a, X
\right]
&=
P(T>t_0\mid A=a,X).
\end{align}
Using the definition $\mu_a(t\mid X)=P(T>t\mid A=a,X)$, we have
\[
P(t_0<T\le t_0+\theta\mid A=a,X)=\mu_a(t_0\mid X)-\mu_a(t_0+\theta\mid X),
\qquad
P(T>t_0\mid A=a,X)=\mu_a(t_0\mid X).
\]
Substituting these into \eqref{eq:dr_score_def} and taking conditional expectations gives
\begin{equation}
\label{eq:Ha_cond_mean}
\mathbb E\!\left[ H_a(O;\theta)\mid A=a,X\right]
=
\Bigl\{\mu_a(t_0\mid X)-\mu_a(t_0+\theta\mid X)\Bigr\}
-\tau\,\mu_a(t_0\mid X)
=
m_a(X;\theta).
\end{equation}

Finally, by definition of the target $\theta_0=q_a(\tau;t_0)$, the corresponding population moment condition is
\begin{equation}
\label{eq:target_moment}
\mathbb E\!\left[m_a(X;\theta_0)\right]=0.
\end{equation}
Because $A$ is a baseline treatment, Assumptions~\ref{ass:consistency} and \ref{ass:exchangeability} imply that the counterfactual survival function under regime $a$ satisfies $P(T_a>t\mid X)=P(T>t\mid A=a,X)=\mu_a(t\mid X)$ for all $t$. Hence, averaging over $X$ yields
$P(t_0<T_a\le t_0+\theta)=E\{\mu_a(t_0\mid X)-\mu_a(t_0+\theta\mid X)\}$ and
$P(T_a>t_0)=E\{\mu_a(t_0\mid X)\}$. Indeed, $\mathbb E[m_a(X;\theta)]=P(t_0<T_a\le t_0+\theta)-\tau P(T_a>t_0)$, so
$\mathbb E[m_a(X;\theta)]=0$ is equivalent to $F_{R_a}(\theta;t_0)=\tau$.

\paragraph{Step 3: Show double robustness.}
We prove that $U^{DR}(\theta_0)=0$ if either (i) the propensity score model is correct
or (ii) the outcome regression model is correct, assuming $G_a$ is correct.

\subparagraph{Case 1: propensity score correct.}
Assume $e^\dagger_a(X)=e_a(X)$ almost surely, while $\mu^\dagger_a$ may be misspecified.
Starting from \eqref{eq:U_DR_pop},
\[
U^{DR}(\theta_0)
=
\mathbb E\!\left[
\frac{I(A=a)}{e_a(X)}\,H_a(O;\theta_0)
\right]
-
\mathbb E\!\left[
\frac{I(A=a)-e_a(X)}{e_a(X)}\,m_a^\dagger(X;\theta_0)
\right].
\]
We analyze the two terms separately.

First, for the augmentation term, apply iterated expectations:
\begin{align*}
\mathbb E\!\left[
\frac{I(A=a)-e_a(X)}{e_a(X)}\,m_a^\dagger(X;\theta_0)
\right]
&=
\mathbb E\!\left[
\frac{m_a^\dagger(X;\theta_0)}{e_a(X)}\,
\mathbb E\!\left\{ I(A=a)-e_a(X)\mid X\right\}
\right]
\\
&=
\mathbb E\!\left[
\frac{m_a^\dagger(X;\theta_0)}{e_a(X)}\,
\{P(A=a\mid X)-e_a(X)\}
\right]
=0.
\end{align*}
Second, for the IW term,
\begin{align*}
\mathbb E\!\left[
\frac{I(A=a)}{e_a(X)}\,H_a(O;\theta_0)
\right]
&=
\mathbb E_X\!\left[
\mathbb E\!\left\{
\frac{I(A=a)}{e_a(X)}\,H_a(O;\theta_0)\mid X
\right\}
\right]
\\
&=
\mathbb E_X\!\left[
\frac{P(A=a\mid X)}{e_a(X)}
\mathbb E\!\left\{H_a(O;\theta_0)\mid A=a,X\right\}
\right]
\\
&=
\mathbb E\!\left[ m_a(X;\theta_0)\right],
\end{align*}
where the last equality uses \eqref{eq:Ha_cond_mean}.
By \eqref{eq:target_moment}, $\mathbb E[m_a(X;\theta_0)]=0$; hence $U^{DR}(\theta_0)=0$.

\subparagraph{Case 2: outcome regression correct.}
Assume $\mu^\dagger_a(t\mid X)=\mu_a(t\mid X)$ for all $t$ (a.s.), hence
$m_a^\dagger(X;\theta)=m_a(X;\theta)$ for all $\theta$ by \eqref{eq:dr_aug_def}--\eqref{eq:dr_aug_limit_def},
while $e^\dagger_a$ may be misspecified.
Use the rearranged form \eqref{eq:U_DR_pop_rearranged} at $\theta=\theta_0$:
\[
U^{DR}(\theta_0)
=
\mathbb E\!\left[
\frac{I(A=a)}{e^\dagger_a(X)}\Bigl\{H_a(O;\theta_0)-m_a(X;\theta_0)\Bigr\}
\right]
+
\mathbb E\!\left[m_a(X;\theta_0)\right].
\]
The second term is $0$ by \eqref{eq:target_moment}.
For the first term, condition on $(A,X)$:
\begin{align*}
\mathbb E\!\left[
\frac{I(A=a)}{e^\dagger_a(X)}\Bigl\{H_a(O;\theta_0)-m_a(X;\theta_0)\Bigr\}
\right]
&=
\mathbb E\!\left[
\frac{I(A=a)}{e^\dagger_a(X)}
\mathbb E\!\left\{
H_a(O;\theta_0)-m_a(X;\theta_0)\mid A,X
\right\}
\right]
\\
&=
\mathbb E\!\left[
\frac{I(A=a)}{e^\dagger_a(X)}
\Bigl(
\mathbb E\!\left\{H_a(O;\theta_0)\mid A=a,X\right\}-m_a(X;\theta_0)
\Bigr)
\right]
\\
&=
\mathbb E\!\left[
\frac{I(A=a)}{e^\dagger_a(X)}\cdot 0
\right]
=0,
\end{align*}
where the last step uses \eqref{eq:Ha_cond_mean}.
Therefore $U^{DR}(\theta_0)=0$ regardless of misspecification of $e^\dagger_a(X)$.

\paragraph{Step 4: Conclude consistency.}
We have shown that, under consistent estimation of the censoring mechanism $G_a$,
the population estimating equation satisfies $U^{DR}(\theta_0)=0$ if either
(i) $e^\dagger_a=e_a$ or (ii) $\mu^\dagger_a=\mu_a$.
Under standard regularity conditions for (possibly nonsmooth) Z-estimation with a unique root
(e.g., monotonicity of the estimating function in $\theta$ and uniform convergence of $U^{DR}_n(\theta)$ to $U^{DR}(\theta)$),
the solution $\widehat q_a^{DR}(\tau;t_0)$ to \eqref{eq:dr_estimating_eq} converges in probability to $\theta_0=q_a(\tau;t_0)$.
This establishes the double robustness claim in Theorem~\ref{thm:dr}. \qed \\

\section{Proof of Theorem \ref{thm:PSQC}: Identification of PSQC Quantile}

We verify identification for each arm $a\in\{0,1\}$ by (i) stating the defining
population equation for $q_a^{PS}$ and (ii) showing it is equivalent to solving
a corresponding observed-data IW moment equation. Throughout, we work under
Assumptions \ref{ass:consistency}--\ref{ass:censoring}, together with
Assumptions \ref{ass:monotonicity}--\ref{ass:principal}. Let
\[
\mathcal{A} := \{S_0=1,S_1=1\},\qquad S_a:=I(T_a>t_0),
\]
and define $\psi(r;\theta):=\tau-I(r\le \theta)$.

\subsection{Step 0 (Definition of the PSQC quantile).}
For each $a\in\{0,1\}$, define $q_a^{PS}$ as the $\tau$-quantile of the residual
lifetime $R_a(t_0)$ in the Always-Survivor stratum $\mathcal{A}$, i.e.,
\begin{equation}
\label{eq:def_q_PSQC}
q_a^{PS}\ \text{is the unique solution }\theta\text{ to}\quad
E\!\left[\psi\!\left(R_a(t_0);\theta\right)\mid \mathcal{A}\right]=0.
\end{equation}
Equivalently (multiplying by $P(\mathcal{A})>0$), $q_a^{PS}$ is the unique root of
\begin{equation}
\label{eq:def_q_PSQC_uncond}
E\!\left[I(\mathcal{A})\,\psi\!\left(R_a(t_0);\theta\right)\right]=0.
\end{equation}

\subsection{Step 1 (Case $a=0$: control PSQC quantile).}
Under Monotonicity (Assumption \ref{ass:monotonicity}), $S_0=1\Rightarrow S_1=1$,
so $\mathcal{A}=\{S_0=1,S_1=1\}=\{S_0=1\}$ almost surely. Hence
\eqref{eq:def_q_PSQC_uncond} for $a=0$ reduces to
\begin{equation}
\label{eq:control_target}
E\!\left[I(S_0=1)\,\psi\!\left(R_0(t_0);\theta\right)\right]=0.
\end{equation}
By Consistency and Exchangeability given $X$
(Assumptions \ref{ass:consistency}--\ref{ass:exchangeability}),
the standard IW identity implies that for any integrable $Z_0$ measurable under arm $0$,
\begin{equation}
\label{eq:ipw_a0}
E[Z_0] \;=\; E\!\left[\frac{I(A=0)}{e_0(X)}\,Z\right],
\qquad e_0(X):=P(A=0\mid X),
\end{equation}
where $Z$ is the observed counterpart of $Z_0$ (under $A=0$).
Applying \eqref{eq:ipw_a0} with
$Z_0 := I(S_0=1)\,\psi(R_0(t_0);\theta)$ yields
\begin{equation}
\label{eq:ipw_control}
E\!\left[I(S_0=1)\,\psi\!\left(R_0(t_0);\theta\right)\right]
=
E\!\left[\frac{I(A=0)}{e_0(X)}\, I(Y>t_0)\,\psi\!\left(R(t_0);\theta\right)\right].
\end{equation}
Finally, by the Independent Censoring assumption \ref{ass:censoring}, the usual IPCW
representation holds (with the censoring survival under $A=0$ denoted
$G_0(\cdot\mid X)$):
\begin{align}
\label{eq:ipcw_control_event}
E\!\left[I(Y>t_0)\,I\!\left(R(t_0)\le \theta\right)\mid A=0,X\right]
&=
E\!\left[\frac{ I(t_0<Y\le t_0+\theta,\ \Delta=1) }
{ G_0(Y\mid X) }\ \Bigg|\ A=0,X\right],\\
\label{eq:ipcw_control_risk}
E\!\left[I(Y>t_0)\mid A=0,X\right]
&=
E\!\left[\frac{ I(Y>t_0) }
{ G_0(t_0\mid X) }\ \Bigg|\ A=0,X\right].
\end{align}
Using $\psi(r;\theta)=\tau-I(r\le \theta)$ and iterated expectation, the left-hand
side of \eqref{eq:ipw_control} equals
\[
E\!\left[
\frac{I(A=0)}{e_0(X)}
\left\{
\frac{ I(t_0<Y\le t_0+\theta,\ \Delta=1) }
{ G_0(Y\mid X) }
-\tau\,
\frac{ I(Y>t_0) }
{ G_0(t_0\mid X) }
\right\}
\right].
\]
Therefore, $q_0^{PS}$ is identified as the unique root of the above IW moment
equation. (This is the special case of \eqref{eq:PSQC_moment_IW} with $a=0$ and
$w_{sel}(A,X)\equiv 1$.)

\subsection{Step 2 (Case $a=1$: treated PSQC quantile).}
We start from the defining equation \eqref{eq:def_q_PSQC} with $a=1$:
\begin{equation}
\label{eq:treat_def}
E\!\left[\psi\!\left(R_1(t_0);\theta\right)\mid \mathcal{A}\right]=0.
\end{equation}
Let $\pi_a(X):=P(S_a=1\mid X)=P(T_a>t_0\mid X)$, $a\in\{0,1\}$. Note that under Monotonicity, $P(\mathcal A\mid S_1=1,X)=P(S_0=1\mid S_1=1,X)=\pi_0(X)/\pi_1(X)$, which motivates the selection weight $w(X)$. By Bayes' rule,
\begin{align}
f_{X\mid \mathcal{A}}(x)
&= \frac{P(\mathcal{A}\mid X=x)f_X(x)}{P(\mathcal{A})}
= \frac{\pi_0(x) f_X(x)}{P(\mathcal{A})}, \nonumber\\
f_{X\mid S_1=1}(x)
&= \frac{P(S_1=1\mid X=x)f_X(x)}{P(S_1=1)}
= \frac{\pi_1(x) f_X(x)}{P(S_1=1)}. \nonumber
\end{align}
Taking the ratio yields the density-ratio identity
\begin{equation}
\label{eq:ratio}
\frac{f_{X\mid \mathcal{A}}(x)}{f_{X\mid S_1=1}(x)}
=
\frac{\pi_0(x)}{\pi_1(x)}\cdot \frac{P(S_1=1)}{P(\mathcal{A})}
=
w(x)\,c,
\qquad
w(x):=\frac{\pi_0(x)}{\pi_1(x)},\ \ c:=\frac{P(S_1=1)}{P(\mathcal{A})}.
\end{equation}
Define $m_\theta(x):=E[\psi(R_1(t_0);\theta)\mid S_1=1,X=x]$.
Then, by change-of-measure using \eqref{eq:ratio},
\begin{align}
E\!\left[\psi(R_1(t_0);\theta)\mid \mathcal{A}\right]
&=
\int m_\theta(x)\,f_{X\mid \mathcal{A}}(x)\,dx \nonumber\\
&=
\int m_\theta(x)\,
\frac{f_{X\mid \mathcal{A}}(x)}{f_{X\mid S_1=1}(x)}\,
f_{X\mid S_1=1}(x)\,dx \nonumber\\
&=
c\,E\!\left[w(X)\,m_\theta(X)\mid S_1=1\right]
=
c\,E\!\left[w(X)\,\psi(R_1(t_0);\theta)\mid S_1=1\right],
\label{eq:change_measure}
\end{align}
where the final equality follows from the tower property. The Principal Ignorability
assumption (Assumption \ref{ass:principal}) ensures that conditioning on $(S_1=1,X)$
is sufficient for the distribution of $R_1(t_0)$ relevant to $\mathcal{A}$ within
treated survivors, so that $m_\theta(\cdot)$ is the correct conditional mean for the
target stratum.

Since $c>0$ does not depend on $\theta$, the unique root of \eqref{eq:treat_def} is
equivalently the unique root of
\begin{equation}
\label{eq:treat_weighted_survivors}
E\!\left[w(X)\,\psi(R_1(t_0);\theta)\mid S_1=1\right]=0.
\end{equation}
Multiplying \eqref{eq:treat_weighted_survivors} by $P(S_1=1)>0$ yields the
unconditional form
\begin{equation}
\label{eq:treat_uncond_survivors}
E\!\left[ I(S_1=1)\,w(X)\,\psi(R_1(t_0);\theta)\right]=0.
\end{equation}

Next, apply the IW identity under $a=1$ (Consistency, Exchangeability, Positivity):
for any integrable $Z_1$ measurable under arm $1$,
\begin{equation}
\label{eq:ipw_a1}
E[Z_1] \;=\; E\!\left[\frac{I(A=1)}{e_1(X)}\,Z\right],
\qquad e_1(X):=P(A=1\mid X),
\end{equation}
with $Z$ the observed counterpart. Taking
$Z_1:=I(S_1=1)\,w(X)\,\psi(R_1(t_0);\theta)$ gives
\begin{equation}
\label{eq:ipw_treat}
E\!\left[ I(S_1=1)\,w(X)\,\psi(R_1(t_0);\theta)\right]
=
E\!\left[\frac{I(A=1)}{e_1(X)}\,I(Y>t_0)\,w(X)\,\psi(R(t_0);\theta)\right].
\end{equation}

Finally, invoke the IPCW identities under arm $1$ (Assumption \ref{ass:censoring})
with censoring survival $G_1(\cdot\mid X)$:
\begin{align}
\label{eq:ipcw_treat_event}
E\!\left[I(Y>t_0)\,I(R(t_0)\le \theta)\mid A=1,X\right]
&=
E\!\left[\frac{ I(t_0<Y\le t_0+\theta,\ \Delta=1) }
{ G_1(Y\mid X) }\ \Bigg|\ A=1,X\right],\\
\label{eq:ipcw_treat_risk}
E\!\left[I(Y>t_0)\mid A=1,X\right]
&=
E\!\left[\frac{ I(Y>t_0) }
{ G_1(t_0\mid X) }\ \Bigg|\ A=1,X\right].
\end{align}
Using $\psi(r;\theta)=\tau-I(r\le \theta)$ and iterated expectation,
\eqref{eq:ipw_treat} becomes
\[
E\!\left[
w(X)\,
\frac{I(A=1)}{e_1(X)}
\left\{
\frac{ I(t_0<Y\le t_0+\theta,\ \Delta=1) }
{ G_1(Y\mid X) }
-\tau\,
\frac{ I(Y>t_0) }
{ G_1(t_0\mid X) }
\right\}
\right]=0.
\]
Recalling that for $A=1$ we have
$w_{sel}(A,X)=w(X)=\pi_0(X)/\pi_1(X)$, the last display is exactly
\eqref{eq:PSQC_moment_IW}. Therefore, the unique solution to
\eqref{eq:PSQC_moment_IW} coincides with $q_1^{PS}$ as defined in
\eqref{eq:def_q_PSQC}.

\subsection{Step 3 (Uniqueness).}
For each $a\in\{0,1\}$, the function
$\theta\mapsto E[\psi(R_a(t_0);\theta)\mid \mathcal{A}]$ is nondecreasing.
Under a mild regularity condition such as continuity of the conditional
distribution function of $R_a(t_0)$ given $\mathcal{A}$ at its $\tau$-quantile,
the root is unique. Each transformation above preserves the root (up to
multiplication by a positive constant), hence the solutions to the corresponding
IW moment equations are unique and equal to $q_a^{PS}$.
\qed

\section{Detailed Data Generation Process}
\label{sec:supp_dgp}

In this section, we provide the full details of the data generation process (DGP) used in the simulation study. The DGP was constructed not only to satisfy the standard causal assumptions (consistency, positivity, exchangeability) but also to strictly enforce the structural assumptions required for the supplementary analysis: \textit{Monotonicity} and \textit{Principal Ignorability}.

\subsection{Covariates and Treatment Assignment}
We generated three baseline covariates $X = (X_{1}, X_{2}, X_{3})^\top$ from a multivariate normal distribution with mean zero, unit variance, and pairwise correlation $\rho = 0.2$:
\begin{equation}
X \sim \mathcal{N}\left( 
\begin{bmatrix} 0 \\ 0 \\ 0 \end{bmatrix}, 
\begin{bmatrix} 
1 & 0.2 & 0.2 \\ 
0.2 & 1 & 0.2 \\ 
0.2 & 0.2 & 1 
\end{bmatrix} 
\right).
\end{equation}
The binary treatment indicator $A$ was generated from a logistic regression model. To assess the double robustness property, we introduced a non-linear quadratic term $X_{1}^2$ into the true propensity score, which serves as the source of model misspecification in scenarios IC (PS Misspecified) and II (Both Misspecified):
\begin{equation}
P(A=1 \mid X) = \text{expit}\left( -0.5 + 0.5 X_{1} + 0.5 X_{2} - 0.2 X_{3} + 1.0 X_{1}^2 \right),
\end{equation}
where $\text{expit}(z) = (1+\exp(-z))^{-1}$.

\subsection{Potential Event Times and Structural Assumptions}
The generation of potential event times $T_0$ and $T_1$ utilized a shared-frailty copula approach to induce correlation between the counterfactuals while satisfying Monotonicity and Principal Ignorability.

Let $\lambda_T(X, a)$ denote the scale parameter for the Weibull distribution of event times under treatment $a$, defined as:
\begin{equation}
\log \lambda_T(X, a) = -1.0 + \beta_{\text{t}} a + 0.5 X_{1} + 0.2 X_{2} + 1.0 X_{1}^2.
\end{equation}
The shape parameter was fixed at $\nu = 1.5$. We set $\beta_{\text{t}} = -0.5$ for the protective scenario (implying treatment reduces the hazard) and $\beta_{\text{t}} = 0$ for the null scenario.

To generate the counterfactuals, we employed a two-stage procedure:

\textbf{Stage 1: Survival to $t_0$ (Enforcing Monotonicity)}
We first drew a latent uniform random variable $U_{\text{pre}} \sim \text{Uniform}(0,1)$, shared across both treatment arms. The potential ``pre-landmark'' event times were calculated as:
\begin{equation}
\tilde{T}_a = \left( \frac{-\log U_{\text{pre}}}{\exp(\log \lambda_T(X, a))} \right)^{1/\nu}, \quad \text{for } a \in \{0,1\}.
\end{equation}
Under the protective scenario ($\beta_{\text{t}} < 0$), we have $\lambda_T(X, 1) < \lambda_T(X, 0)$, which implies $\tilde{T}_1 > \tilde{T}_0$ for a fixed $U_{\text{pre}}$. Consequently, if an individual survives to $t_0$ under control ($\tilde{T}_0 > t_0$), they necessarily survive under treatment ($\tilde{T}_1 > t_0$), thereby satisfying the Monotonicity assumption ($S_1 \ge S_0$).

\textbf{Stage 2: Residual Lifetime (Enforcing Principal Ignorability)}
If the generated pre-time $\tilde{T}_a$ exceeded $t_0$, the actual event time was constructed using a \emph{new}, independent random draw $U_{\text{post}} \sim \text{Uniform}(0,1)$. The potential outcome $T_a$ was defined as:
\begin{equation}
T_a = 
\begin{cases} 
\tilde{T}_a & \text{if } \tilde{T}_a \le t_0 \quad (\text{Dies before } t_0), \\
t_0 + \left( \frac{-\log U_{\text{post}}}{\exp(\log \lambda_T(X, a))} \right)^{1/\nu} & \text{if } \tilde{T}_a > t_0 \quad (\text{Survives to } t_0).
\end{cases}
\end{equation}
By using fresh random noise ($U_{\text{post}, a}$) for the residual component, we ensured that the residual lifetime $R_a(t_0) = T_a - t_0$ is conditionally independent of the latent survival status determined by $U_{\text{pre}}$. This construction implies that, conditional on $X$, the post-$t_0$ residual component is generated independently of the pre-$t_0$ survival indicator. In particular, it is designed so that the \emph{principal ignorability} condition used for identifying the PSQC holds, namely that for all measurable $h$,
\[
E\!\left[h\!\left(R_1(t_0)\right)\mid S_0=1,S_1=1,X\right]
=
E\!\left[h\!\left(R_1(t_0)\right)\mid S_1=1,X\right].
\]
The observed event time was then set to $T = T_A$ according to the Consistency assumption.

\subsection{Censoring Mechanism}
Censoring times $C$ were generated from an exponential distribution with rate parameter $\lambda_C(X, A)$, creating informative censoring dependent on both treatment and covariates:
\begin{equation}
C \sim \text{Exponential}\left( \exp\left( -0.5 - 1.0 A + 0.1 X_{3} + 0.1 X_{1} X_{3} \right) \right).
\end{equation}
This specification creates differential censoring rates between arms and induces dependence on the interaction term $X_{1}X_{3}$, which must be correctly modeled in the IPCW weights to avoid selection bias. The observed follow-up time was defined as $Y = \min(T, C)$ with event indicator $\Delta = I(T \le C)$.

\subsection{Benchmark Scenario: Violation of Structural Assumptions}
To further justify our primary focus on the OSQC, we also considered an ``ordinary'' data generation process where the strict structural assumptions of Monotonicity and Principal Ignorability were \emph{not} enforced. In this simple setup (denoted as \texttt{generate\_data\_alt} in the simulation code), potential event times $T_0$ and $T_1$ were generated using independent random draws conditional on covariates, rather than the shared-frailty approach using $U_{\text{pre}},U_{\text{post}}\sim \text{Uniform}(0,1)$ described in Section \ref{sec:supp_dgp}.

Under this standard conditionally independent structure:
\begin{enumerate}
    \item The assumptions required for identifying the OSQC (Consistency, Positivity, Exchangeability, Independent Censoring) remain valid. Consequently, the proposed IW and DR estimators remain consistent and perform perfectly.
    \item The assumptions required for identifying the PSQC (Monotonicity, Principal Ignorability) are violated. Specifically, the independence of residuals $R_a$ from survivor status $S_a$ no longer holds in the reweighted pseudo-population, causing the PSQC estimator to yield biased results.
\end{enumerate}

This contrast highlights a critical practical distinction: the OSQC is a robust estimand identifiable under standard causal assumptions, whereas the PSQC relies on fragile, untestable structural constraints. This observation reinforces our decision to prioritize the OSQC as the primary clinical estimand, utilizing the PSQC framework strictly as a supplementary analysis tool.

\section{Additional Simulation Results}
\label{sec:supp_sim}

\begin{table}[ht]
\centering
\caption{Simulation results for landmark time $t_0=0.3$, reporting Bias, Standard Error (SE), and Wald-type 95\% Coverage Probability (CP). Bias is calculated relative to the true OSQC for KM, IW, DR and true PSQC for PS, where true values were calculated using $10^7$ Monte Carlo simulations. Note that results of the KM estimator are the same across all scenarios. \textbf{Abbreviations:} CC, both propensity and outcome models correct; CI, correct propensity but misspecified outcome; IC, correct outcome but misspecified propensity; II, both misspecified.}\label{tab:sim_results_t0.3}
\begin{tabular}{ll | ccc ccc | ccc ccc}
\hline\hline
\multicolumn{2}{l|}{\multirow{2}{*}{\textbf{\shortstack{Landmark \\ $t_{0}=0.3$}}}} & \multicolumn{6}{c|}{$\tau=0.3$} & \multicolumn{6}{c}{$\tau=0.5$} \\
\multicolumn{2}{l|}{} & \multicolumn{3}{c}{\textbf{$N=500$}} & \multicolumn{3}{c|}{$N=2000$} & \multicolumn{3}{c}{\textbf{$N=500$}} & \multicolumn{3}{c}{$N=2000$} \\  
\cmidrule(lr){3-5} \cmidrule(lr){6-8} \cmidrule(lr){9-11} \cmidrule(lr){12-14}
\multicolumn{2}{l|}{\textbf{\shortstack{Scenario \\ $\&$Method}}} & \textbf{Bias} & \textbf{SE} & \textbf{CP} & \textbf{Bias} & \textbf{SE} & \textbf{CP} & \textbf{Bias} & \textbf{SE} & \textbf{CP} & \textbf{Bias} & \textbf{SE} & \textbf{CP} \\ \hline
\multicolumn{2}{c}{\textit{ $\beta_{\text{t}} = 0$}} & \multicolumn{6}{c}{\textit{OSQC = PSQC = 0}} & \multicolumn{6}{c}{\textit{OSQC = PSQC = 0}} \\
\hline
\multirow{4}{*}{CC}
 & KM & -0.45 & 0.27 & 0.66 & -0.43 & 0.13 & 0.08 & -0.34 & 0.16 & 0.51 & -0.34 & 0.08 & 0.01 \\
 & IW & -0.00 & 0.13 & 0.94 & -0.00 & 0.06 & 0.96 & 0.00 & 0.19 & 0.95 & -0.01 & 0.09 & 0.94 \\
 & DR & -0.00 & 0.13 & 0.94 & -0.00 & 0.06 & 0.96 & -0.00 & 0.19 & 0.95 & -0.01 & 0.09 & 0.95 \\
 & PS & -0.00 & 0.13 & 0.94 & -0.00 & 0.06 & 0.96 & -0.00 & 0.19 & 0.95 & -0.01 & 0.09 & 0.95 \\
\hline
\multirow{3}{*}{CI}
 & IW & -0.00 & 0.13 & 0.94 & -0.00 & 0.06 & 0.96 & 0.00 & 0.19 & 0.95 & -0.01 & 0.09 & 0.94 \\
 & DR & -0.01 & 0.17 & 0.94 & -0.01 & 0.11 & 0.96 & 0.00 & 0.22 & 0.95 & -0.01 & 0.14 & 0.95 \\
 & PS & -0.02 & 0.13 & 0.94 & -0.02 & 0.06 & 0.95 & -0.02 & 0.19 & 0.96 & -0.03 & 0.09 & 0.94 \\
\hline
\multirow{3}{*}{IC}
 & IW & -0.22 & 0.12 & 0.57 & -0.22 & 0.06 & 0.03 & -0.30 & 0.18 & 0.66 & -0.30 & 0.09 & 0.06 \\
 & DR & 0.00 & 0.12 & 0.95 & -0.00 & 0.05 & 0.96 & 0.00 & 0.17 & 0.94 & -0.01 & 0.08 & 0.95 \\
 & PS & -0.22 & 0.12 & 0.58 & -0.22 & 0.06 & 0.03 & -0.30 & 0.18 & 0.66 & -0.30 & 0.09 & 0.06 \\
\hline
\multirow{3}{*}{II}
 & IW & -0.22 & 0.12 & 0.57 & -0.22 & 0.06 & 0.03 & -0.30 & 0.18 & 0.66 & -0.30 & 0.09 & 0.06 \\
 & DR & -0.22 & 0.12 & 0.58 & -0.22 & 0.06 & 0.03 & -0.31 & 0.18 & 0.62 & -0.31 & 0.09 & 0.05 \\
 & PS & -0.23 & 0.12 & 0.55 & -0.23 & 0.06 & 0.01 & -0.32 & 0.18 & 0.61 & -0.32 & 0.09 & 0.04 \\
\hline
\hline
\multicolumn{2}{c}{\textit{ $\beta_{\text{t}} = -0.5$}} & \multicolumn{6}{c}{\textit{OSQC $\approx$ 0.20, \quad PSQC $\approx$ 0.25}} & \multicolumn{6}{c}{\textit{OSQC $\approx$ 0.37, \quad PSQC $\approx$ 0.43}} \\
\hline
\multirow{4}{*}{CC}
 & KM & -0.15 & 0.29 & 0.94 & -0.13 & 0.14 & 0.86 & -0.42 & 0.18 & 0.41 & -0.42 & 0.09 & 0.00 \\
 & IW & 0.00 & 0.15 & 0.94 & -0.00 & 0.07 & 0.96 & 0.01 & 0.21 & 0.95 & -0.01 & 0.10 & 0.95 \\
 & DR & 0.00 & 0.14 & 0.94 & -0.00 & 0.07 & 0.96 & 0.00 & 0.21 & 0.95 & -0.01 & 0.10 & 0.95 \\
 & PS & 0.00 & 0.15 & 0.95 & -0.00 & 0.07 & 0.96 & 0.01 & 0.21 & 0.96 & -0.00 & 0.11 & 0.95 \\
\hline
\multirow{3}{*}{CI}
 & IW & 0.00 & 0.15 & 0.94 & -0.00 & 0.07 & 0.96 & 0.01 & 0.21 & 0.95 & -0.01 & 0.10 & 0.95 \\
 & DR & -0.01 & 0.18 & 0.95 & -0.01 & 0.11 & 0.96 & 0.01 & 0.24 & 0.96 & -0.00 & 0.14 & 0.95 \\
 & PS & -0.06 & 0.14 & 0.93 & -0.07 & 0.07 & 0.85 & -0.07 & 0.21 & 0.95 & -0.08 & 0.10 & 0.88 \\
\hline
\multirow{3}{*}{IC}
 & IW & -0.27 & 0.13 & 0.52 & -0.27 & 0.06 & 0.01 & -0.37 & 0.20 & 0.57 & -0.37 & 0.10 & 0.03 \\
 & DR & 0.01 & 0.13 & 0.94 & -0.00 & 0.06 & 0.96 & 0.01 & 0.20 & 0.95 & -0.00 & 0.09 & 0.95 \\
 & PS & -0.26 & 0.14 & 0.55 & -0.25 & 0.06 & 0.02 & -0.35 & 0.20 & 0.64 & -0.35 & 0.10 & 0.06 \\
\hline
\multirow{3}{*}{II}
 & IW & -0.27 & 0.13 & 0.52 & -0.27 & 0.06 & 0.01 & -0.37 & 0.20 & 0.57 & -0.37 & 0.10 & 0.03 \\
 & DR & -0.28 & 0.13 & 0.50 & -0.27 & 0.06 & 0.01 & -0.39 & 0.20 & 0.53 & -0.39 & 0.10 & 0.02 \\
 & PS & -0.33 & 0.13 & 0.33 & -0.33 & 0.06 & 0.00 & -0.44 & 0.20 & 0.45 & -0.44 & 0.10 & 0.01 \\
\hline\hline
\end{tabular}
\end{table}

\begin{table}[ht]
\centering
\caption{Simulation results for landmark time $t_0=0.7$, reporting Bias, Standard Error (SE), and Wald-type 95\% Coverage Probability (CP). Bias is calculated relative to the true OSQC for KM, IW, DR and true PSQC for PS, where true values were calculated using $10^7$ Monte Carlo simulations. Note that results of the KM estimator are the same across all scenarios. \textbf{Abbreviations:} CC, both propensity and outcome models correct; CI, correct propensity but misspecified outcome; IC, correct outcome but misspecified propensity; II, both misspecified.}\label{tab:sim_results_t0.7}
\begin{tabular}{ll | ccc  ccc | ccc ccc}
\hline\hline
\multicolumn{2}{l|}{\multirow{2}{*}{\textbf{\shortstack{Landmark \\ $t_{0}=0.7$}}}} & \multicolumn{6}{c|}{$\tau=0.3$} & \multicolumn{6}{c}{$\tau=0.5$} \\
\multicolumn{2}{l|}{} & \multicolumn{3}{c}{\textbf{$N=500$}} & \multicolumn{3}{c|}{$N=2000$} & \multicolumn{3}{c}{\textbf{$N=500$}} & \multicolumn{3}{c}{$N=2000$} \\
\cmidrule(lr){3-5} \cmidrule(lr){6-8} \cmidrule(lr){9-11} \cmidrule(lr){12-14}
\multicolumn{2}{l|}{\textbf{\shortstack{Scenario \\ $\&$Method}}} & \textbf{Bias} & \textbf{SE} & \textbf{CP} & \textbf{Bias} & \textbf{SE} & \textbf{CP} & \textbf{Bias} & \textbf{SE} & \textbf{CP} & \textbf{Bias} & \textbf{SE} & \textbf{CP} \\ \hline

\multicolumn{2}{c}{\textit{ $\beta_{\text{t}} = 0$}} & \multicolumn{6}{c}{\textit{OSQC = PSQC = 0}} & \multicolumn{6}{c}{\textit{OSQC = PSQC = 0}} \\
\hline
\multirow{4}{*}{CC} & KM  & -0.35 & 0.34 & 0.90 & -0.32 & 0.16 & 0.57 & -0.25 & 0.22 & 0.87 & -0.25 & 0.11 & 0.38 \\
 & IW  & -0.01 & 0.17 & 0.96 & -0.01 & 0.08 & 0.95 & -0.01 & 0.25 & 0.95 & -0.01 & 0.12 & 0.96 \\
 & DR  & -0.01 & 0.17 & 0.95 & -0.01 & 0.08 & 0.96 & -0.02 & 0.25 & 0.96 & -0.01 & 0.12 & 0.96 \\
 & PS  & -0.01 & 0.17 & 0.95 & -0.01 & 0.08 & 0.95 & -0.02 & 0.25 & 0.96 & -0.01 & 0.12 & 0.96 \\
\hline
\multirow{3}{*}{CI} & IW  & -0.01 & 0.17 & 0.96 & -0.01 & 0.08 & 0.95 & -0.01 & 0.25 & 0.95 & -0.01 & 0.12 & 0.96 \\
 & DR  & -0.01 & 0.20 & 0.96 & -0.01 & 0.10 & 0.96 & -0.03 & 0.29 & 0.95 & -0.03 & 0.15 & 0.96 \\
 & PS  & -0.03 & 0.17 & 0.95 & -0.03 & 0.08 & 0.95 & -0.04 & 0.25 & 0.95 & -0.03 & 0.12 & 0.96 \\
\hline
\multirow{3}{*}{IC} & IW  & -0.16 & 0.16 & 0.83 & -0.15 & 0.08 & 0.52 & -0.20 & 0.24 & 0.85 & -0.20 & 0.12 & 0.64 \\
 & DR  & -0.01 & 0.15 & 0.95 & -0.01 & 0.07 & 0.96 & -0.02 & 0.24 & 0.95 & -0.01 & 0.12 & 0.96 \\
 & PS  & -0.16 & 0.16 & 0.83 & -0.15 & 0.08 & 0.52 & -0.20 & 0.24 & 0.85 & -0.20 & 0.12 & 0.64 \\
\hline
\multirow{3}{*}{II} & IW  & -0.16 & 0.16 & 0.83 & -0.15 & 0.08 & 0.52 & -0.20 & 0.24 & 0.85 & -0.20 & 0.12 & 0.64 \\
 & DR  & -0.17 & 0.16 & 0.82 & -0.16 & 0.08 & 0.46 & -0.22 & 0.24 & 0.83 & -0.22 & 0.12 & 0.56 \\
 & PS  & -0.17 & 0.16 & 0.82 & -0.17 & 0.08 & 0.44 & -0.22 & 0.24 & 0.83 & -0.22 & 0.12 & 0.55 \\
\hline
 \multicolumn{2}{c}{\textit{ $\beta_{\text{t}} = -0.5$}} & \multicolumn{6}{c}{\textit{OSQC $\approx$ 0.23, \quad PSQC $\approx$ 0.29}} & \multicolumn{6}{c}{\textit{OSQC $\approx$ 0.40, \quad PSQC $\approx$ 0.47}} \\
\hline
\multirow{4}{*}{CC} & KM  & 0.03 & 0.36 & 0.95 & 0.03 & 0.17 & 0.94 & -0.32 & 0.23 & 0.78 & -0.32 & 0.11 & 0.42 \\
 & IW  & -0.01 & 0.18 & 0.96 & -0.01 & 0.09 & 0.96 & -0.02 & 0.28 & 0.97 & -0.01 & 0.14 & 0.96 \\
 & DR  & -0.02 & 0.18 & 0.97 & -0.02 & 0.09 & 0.96 & -0.02 & 0.28 & 0.96 & -0.02 & 0.14 & 0.96 \\
 & PS  & -0.02 & 0.18 & 0.97 & -0.02 & 0.09 & 0.96 & -0.02 & 0.29 & 0.96 & -0.02 & 0.14 & 0.96 \\
\hline
\multirow{3}{*}{CI} & IW  & -0.01 & 0.18 & 0.96 & -0.01 & 0.09 & 0.96 & -0.02 & 0.28 & 0.97 & -0.01 & 0.14 & 0.96 \\
 & DR  & -0.03 & 0.20 & 0.97 & -0.03 & 0.10 & 0.97 & -0.03 & 0.31 & 0.97 & -0.03 & 0.17 & 0.96 \\
 & PS  & -0.10 & 0.18 & 0.94 & -0.11 & 0.09 & 0.84 & -0.09 & 0.29 & 0.96 & -0.10 & 0.14 & 0.87 \\
\hline
\multirow{3}{*}{IC} & IW  & -0.20 & 0.17 & 0.76 & -0.19 & 0.08 & 0.41 & -0.25 & 0.25 & 0.78 & -0.25 & 0.12 & 0.55 \\
 & DR  & -0.01 & 0.17 & 0.96 & -0.01 & 0.08 & 0.96 & -0.02 & 0.25 & 0.96 & -0.01 & 0.12 & 0.96 \\
 & PS  & -0.19 & 0.17 & 0.77 & -0.19 & 0.08 & 0.46 & -0.23 & 0.25 & 0.82 & -0.23 & 0.12 & 0.63 \\
\hline
\multirow{3}{*}{II} & IW  & -0.20 & 0.17 & 0.76 & -0.19 & 0.08 & 0.41 & -0.25 & 0.25 & 0.78 & -0.25 & 0.12 & 0.55 \\
 & DR  & -0.21 & 0.17 & 0.75 & -0.20 & 0.08 & 0.35 & -0.28 & 0.25 & 0.75 & -0.28 & 0.12 & 0.46 \\
 & PS  & -0.26 & 0.17 & 0.65 & -0.26 & 0.08 & 0.19 & -0.33 & 0.25 & 0.69 & -0.33 & 0.12 & 0.31 \\
\hline\hline
\end{tabular}
\end{table}

Table~\ref{tab:sim_results_t0.3} and Table \ref{tab:sim_results_t0.7} summarize the simulation results for the alternative landmark time $t_0\in\{0.3,0.7\}$, which exhibit patterns fully consistent with those observed at $t_0=0.5$ in the main manuscript. As anticipated, the KM estimator displays persistent bias across all scenarios due to its inability to adjust for confounding, with limited improvement in coverage as the sample size increases. The IW estimator performs well only when the propensity score model is correctly specified (\textbf{CC} and \textbf{CI}), but incurs substantial bias under propensity misspecification (\textbf{IC} and \textbf{II}). In contrast, the proposed DR estimator continues to demonstrate its double-robust property, remaining essentially unbiased across all robust scenarios (\textbf{CC}, \textbf{CI}, and \textbf{IC}), with bias uniformly close to zero and empirical coverage probabilities approaching the nominal 95\% level as $N$ increases. When both nuisance models are misspecified (\textbf{II}), all estimators exhibit non-negligible bias, as theoretically expected. Finally, under a protective treatment effect ($\beta_t=-0.5$), the divergence between the OSQC and the PSQC becomes evident: while the DR estimator continues to reliably target the OSQC under partial misspecification, the PS estimator accurately recovers the latent PSQC only under full correct specification (\textbf{CC}) and displays pronounced bias otherwise.

\end{document}